\documentclass[fleqn,usenatbib]{mnras}

\usepackage[T1]{fontenc}

\DeclareRobustCommand{\VAN}[3]{#2}
\let\VANthebibliography\thebibliography
\def\thebibliography{\DeclareRobustCommand{\VAN}[3]{##3}\VANthebibliography}

\usepackage{graphicx}	
\usepackage{amsmath}	
\usepackage{anyfontsize} 
\usepackage{newtxtext,newtxmath}
\usepackage{physics}
\usepackage{comment}

\title[Search for Ly$\alpha$ nebulae in the HSC joint field]{RIDEN pilot survey: broad-band selection of candidate quasars with extended Lyman-$\alpha$ nebulae using CLAUDS--HSC-SSP--DUNES$^2$ joint data}

\author[R. Shimakawa]{Rhythm Shimakawa,$^{1,2}$\thanks{E-mail: rhythm.shimakawa@aoni.waseda.jp}
Satoshi Kikuta,$^3$
Haruka Kusakabe,$^{4,5}$
Marcin Sawicki,$^6$
Yongming Liang,$^4$
\newauthor
Rieko Momose,$^7$
Stephen Gwyn$^8$
and Guillaume Desprez$^{6,9}$
\\
$^{1}$Waseda Institute for Advanced Study (WIAS), Waseda University, 1-21-1, Nishi-Waseda, Shinjuku, Tokyo 169-0051, Japan\\
$^{2}$Center for Data Science, Waseda University, 1-6-1, Nishi-Waseda, Shinjuku, Tokyo 169-0051, Japan\\
$^{3}$Department of Astronomy, School of Science, The University of Tokyo, 7-3-1 Hongo, Bunkyo-ku, Tokyo 113-0033, Japan\\
$^{4}$National Astronomical Observatory of Japan (NAOJ), National Institutes of Natural Sciences, Osawa, Mitaka, Tokyo 181-8588, Japan\\
$^{5}$Department of General Systems Studies, Graduate School of Arts and Sciences, The University of Tokyo, 3-8-1 Komaba, Meguro-ku, Tokyo, 153-8902, Japan\\
$^{6}$Institute for Computational Astrophysics and Deparment of Astronomy and Physics, Saint Mary’s University, 923 Robie Street, Halifax, NS B3H 3C3, Canada\\
$^{7}$Carnegie Observatories, Carnegie Science, 813 Santa Barbara Street, Pasadena, CA 91101, USA\\
$^{8}$National Research Council of Canada, Herzberg Astronomy and Astrophysics Research Centre, 5077 West Saanich Road, Victoria, BC V9E 2E7, Canada\\
$^{9}$Kapteyn Astronomical Institute, University of Groningen, P.O. Box 800, 9700AV Groningen, The Netherlands
}

\date{Accepted 2025 June 4. Received 2025 May 8; in original form 2024 November 28}

\pubyear{2025}

\begin{document}
\label{firstpage}
\pagerange{\pageref{firstpage}--\pageref{lastpage}}
\maketitle

\begin{abstract}
The Vera C. Rubin Observatory will conduct the Legacy Survey of Space and Time (LSST), delivering deep, multi-band ($ugrizy$) imaging data across 18,000 square degrees over the next decade.
Before this ultra-wide-field survey, we constructed a broad-band Ly$\alpha$ imaging toward 483 SDSS/BOSS quasars at $z=$ 1.9--3.0, using deep, wide-field ultraviolet to near-infrared ($u$-to-$K$) data from the Hyper Suprime-Cam Subaru Strategic Survey (HSC-SSP), the CFHT Large Area U-band Deep Survey (CLAUDS), the Deep UKIRT Near-Infrared Steward Survey (DUNES$^2$), and additional public data covering 13 square degrees.
Our broad-band selection allowed us to select 24 candidate quasar nebulae that exhibit $u$ or $g$ band excess over 50--170 kpc, some of which exhibit asymmetrical extended features similar to those seen in previously discovered giant nebulae.
We then investigated whether the Ly$\alpha$ morphology of quasar nebulae differs between two redshift intervals, $z=$ 1.9--2.3 and $z=$ 2.3--3.0, and examined environmental dependence based on a control sample.
Comparison results show no significant difference in asymmetry within Ly$\alpha$ nebulae between the two redshift intervals.
Furthermore, we found no systematic differences in overdensities around the complete quasar samples, quasars with large Ly$\alpha$ nebulae, and control samples, while the most extended nebula appears to be located in the high-density region.
Further verification analyses are required since the current dataset lacks spectroscopic confirmation for both quasar nebulae and their surrounding neighbours.
Nevertheless, the results demonstrate the great potential of the Rubin LSST to discover giant Ly$\alpha$ nebulae on an unprecedented scale. 
\end{abstract}





\section{Introduction}
\label{s1}

Lyman-$\alpha$ (Ly$\alpha$) emission has long been used as a tracer of high-redshift objects in the optical regime, which can be powered by such as hot stars, black holes, and accreting cool gas, and can be observed through radiative transfer (e.g., \citealt{Partridge1967,Dijkstra2017}, and references therein).
In particular, extended Ly$\alpha$ nebulae around luminous quasars are useful test beds for understanding spatial properties and dynamics of cool gas on the scales of the circumgalactic medium (CGM) and the intergalactic medium (IGM) around supermassive black holes and host galaxies (e.g., \citealt{McCarthy1987,McCarthy1993,vanOjik1997,Villar-Martin2007,Faucher-Giguere2010,Goerdt2010,Tumlinson2017,Cantalupo2017,Kimock2021}, and references therein).
Enormous Ly$\alpha$ nebula (ELAN) is the most notable case among them, defined as Ly$\alpha$ nebulosity with particularly high surface brightness spreading beyond hundreds of kpc \citep{Cantalupo2014,Hennawi2015,Cai2017,Cai2018,ArrigoniBattaia2018,Li2024}. 
ELAN is often considered a vast gas reservoir surrounding a massive halo as they are found in protocluster regions; hence, they provide unique insights into the gas-feeding mechanism in massive haloes at high redshifts \citep{ArrigoniBattaia2018,Chen2021,Vayner2023,Zhang2025b}.

Expanding the ELANe sample is therefore crucial for a comprehensive understanding of CGM and IGM associated with high-$z$ massive haloes.
However, presently, not many cases of such prominently extended Ly$\alpha$ nebulae have been discovered (specific figures depend on the definition).
One reason is that deep observations with integral-field-unit (IFU) or narrow-band are necessary to identify them  \citep{Matsuda2011,Borisova2016,ArrigoniBattaia2019,Cai2019,Farina2019,Drake2019,Kikuta2019,O'Sullivan2020,Fossati2021,Li2024,ColomaPuga2025}. 
IFU survey is generally more efficient in achieving depth, whereas narrow-band imaging is better suited for covering a wider field-of-view.
In either case, it is challenging to carry out a systematic deep exploration when considering the observational costs.
Another reason is that ELANs are extremely rare: \citet{ArrigoniBattaia2019} have suggested the abundance of ELANe among luminous quasars with $-28.29\leq M_{1450}\leq-25.65$ is $\sim1\%$ or less, meaning that the pruning of quasar samples is vital to increase identifications of ELANe within reasonable observing time (see also \citealt{Shimakawa2022,Li2024}).

To overcome these limitations, \citet{Shimakawa2022} tested a systematic search for the extended Ly$\alpha$ nebulae associated with quasars at $z>2$ using the advanced wide-field imaging data delivered by the Hyper Suprime-Cam Subaru Strategic Program on the 8.2 m Subaru Telescope (HSC-SSP; \citealt{Aihara2018,Miyazaki2018,Furusawa2018,Kawanomoto2018,Komiyama2018}). 
The HSC-SSP surveyed approximately 1,200 deg$^2$ ($\sim$1,000 deg$^2$ as of the public data release 3; \citealt{Aihara2022}) with the five broad-band filters ($grizy$) down to, e.g., $\gtrsim26$ mag in $5\sigma$ limiting magnitude with seeing full-width-half-maximum (FWHM) of $\sim0.6$ arcsec in the $i$-band\footnote{\url{https://hsc.mtk.nao.ac.jp/ssp/data-release}}. 
\citet{Shimakawa2022} searched extended Ly$\alpha$ nebulae around 8,683 SDSS-IV/eBOSS quasars at $z=$ 2.3--3.0 \citep{Myers2015,Lyke2020} using three broad-band imaging data ($gri$) to extract Ly$\alpha$ emission buried in the $g$-band. 
\citet{Prescott2012,Prescott2013} had previously applied a similar approach to the 9.4 deg$^2$ Bo\"otes field with two broad-band filters and confirmed that it worked successfully.
Although the broad-band technique cannot obtain Ly$\alpha$ flux and surface brightness with high precision due to such as IGM absorption effects and flux contributions from other UV emission lines, it has the great advantage of being able to explore Ly$\alpha$ nebulae over overpoweringly wide field without any additional cost under the legacy surveys and to provide us with promising ELAN candidates for follow-up in narrow-band and integral-field unit surveys.
It also allows us to address controversial topics regarding systematic properties of Ly$\alpha$ nebulae, such as the number density of ELANe and environmental dependence on the size of Ly$\alpha$ nebulae, with unrivalled datasets.
Even a blind search across the entire survey field is technically feasible, whereas it is out of the scope of this paper.
Furthermore, we can apply this technique to the upcoming Legacy Survey of Space and Time (LSST) on the Vera C. Rubin Observatory \citep{Ivezic2019}, enabling unprecedentedly wide-field survey of Ly$\alpha$ nebulae at $z\gtrsim2$ over 18,000 deg$^2$ of the southern sky. 

With these backgrounds, this study tests a pilot survey toward the forthcoming Legacy Survey of Space and Time (LSST) on the Rubin Observatory, termed Rubin Image Detection for Enormous Nebulae (RIDEN\footnote{Inspired by one of the greatest sumo wrestlers (Raiden Tameemon, 1767-1825). He was 1.97 meters tall, by far the tallest at that time.}), based on the results of the previous methodology analyses \citep{Shimakawa2022}.
We employ an internal dataset termed HSC Joint-Data (HSC joint collaboration, in preparation), consisting of the Deep and UltraDeep layers in the Second Public Data Release of the Hyper Suprime-Cam Subaru Strategic Program (HSC-SSP PDR2; \citealt{Aihara2019}), the CFHT Large Area U-band Deep Survey (CLAUDS; \citealt{Sawicki2019})\footnote{\url{https://www.clauds.net}}, the Deep UKIRT Near-Infrared Steward Survey (DUNES$^2$; Egami et al. in preparation), and some additional public data.
Several scientific results have already been reported using this large dataset, including discoveries of galaxy clusters at $z>2$ \citep{Kiyota2025}, and environmental dependence in high-redshift protoclusters \citep{Toshikawa2024}, and derivations of star-formation histories of early-type galaxies \citep{Ali2024}.
The HSC Joint-Data cover four survey fields, E-COSMOS, DEEP2-3, ELAIS-N1, and SXDS+XMM-LSS, which amounts to a total area of 13 deg$^2$, covering 483 quasars at $z=$ 1.9--3.0 (Section~\ref{s2}).

We search for Ly$\alpha$ nebulae using this multi-band dataset.
While this study covers a smaller area than our previous study \citep{Shimakawa2022}, it has several advantages owing to additional imaging data at the $u$-band and the near-infrared (NIR) bands.
In particular, $u$ ($u^\ast$)-band data enables us to search for Ly$\alpha$ nebulae at $z=$ 1.9--2.3, which can go deeper than that using the $g$-band ($z=$ 2.3--3.0) owing to two times narrower filter width, i.e., two times greater line sensitivity in the same imaging depth, and also $\sim2.5$ times brighter surface brightness given the redshift dimming $\propto(1+z)^4$.
Furthermore, improved photometric redshift estimations based on multi-band photometry from UV to NIR help us to mask foreground and background contaminants when we establish Ly$\alpha$ line images from the broad-band data (Section~\ref{s3}); they also help search for neighbour galaxies associated with the Ly$\alpha$ nebulae.
We constrain spatial extents and radial profiles of the selected Ly$\alpha$ nebulae to check for significantly extended nebulae in our sample (Section~\ref{s5}).
We then discuss their Ly$\alpha$ morphology and environmental dependence by comparing them with quasars without giant Ly$\alpha$ nebulae and control sample (Section~\ref{s5}).
Lastly, we summarise the results obtained and discuss them in Section~\ref{s6}.

We assume cosmological parameters of $\Omega_M=0.310$, $\Omega_\Lambda=0.689$, and $H_0=67.7$ km~s$^{-1}$Mpc$^{-1}$ in a flat Lambda cold dark matter model, which are consistent with those from the Planck 2018 VI results \citep{PlanckCollaboration2020}. 
Additionally, we adopt the AB magnitude system \citep{Oke1983} throughout the paper. 
When we refer to the figures and tables included in this paper, we use capitalised words (e.g., Figure~1 or Table~1), making them easily distinguishable from those in the literature (e.g., figure~1 or table~1).


\section{Data and target selection}
\label{s2}

\begin{figure}
\centering
\includegraphics[width=7.5cm]{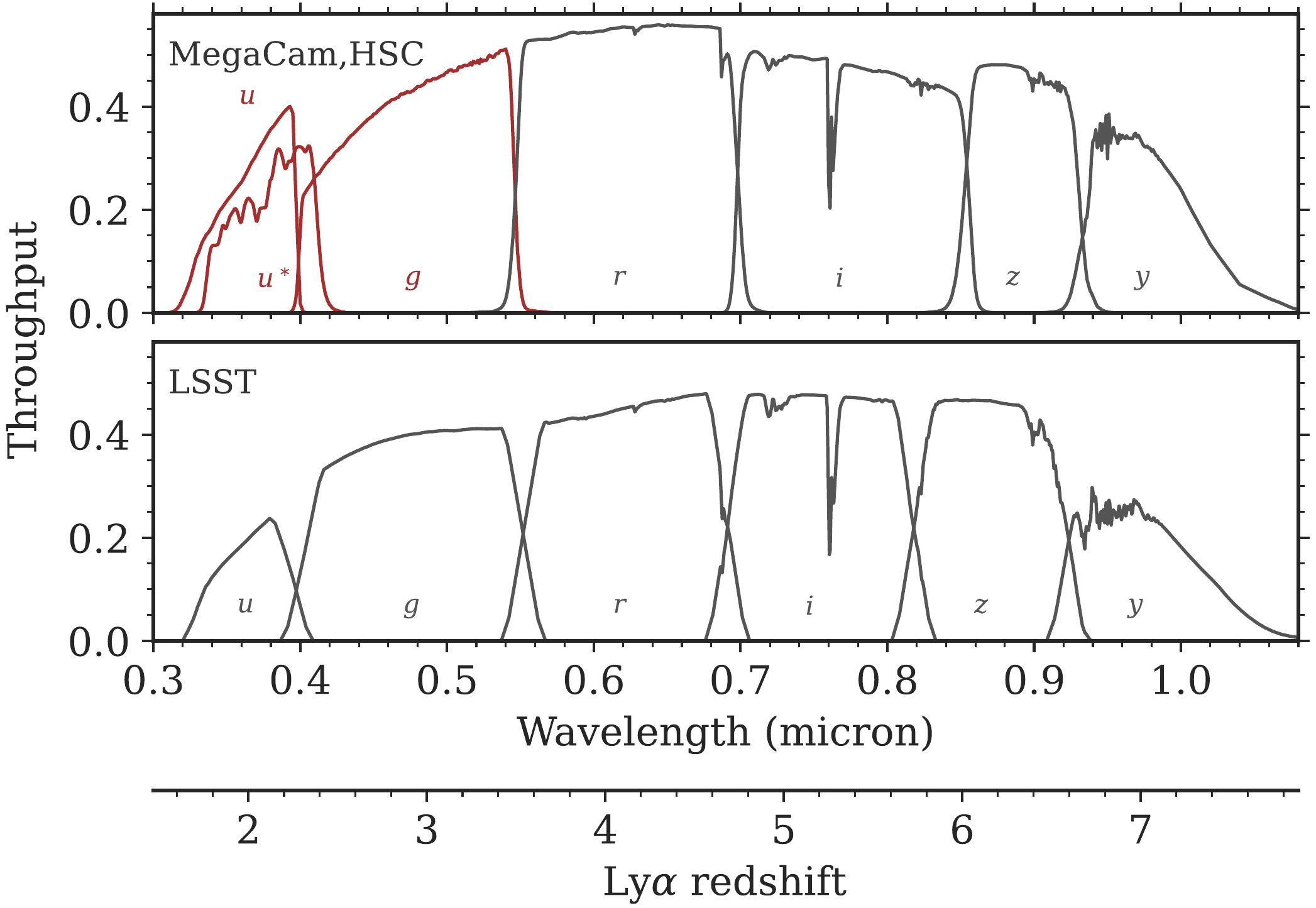}
\caption{
Effective throughput of broad-band filters ($ugrizy$) on Megacam and HSC (top) and LSST Camera (bottom). 
Ly$\alpha$ emissions from our target quasars at $z=$ 1.9--3.0 can be traced using $u$, $u^\ast$, or $g$-band filters.
}
\label{fig1}
\end{figure}

We adopt 483 quasars at $z=$ 1.9--3.0 from the DR16Q catalogue \citep{Lyke2020} covered by the HSC-SSP--CLAUDS--DUNES$^2$ joint field (HCD-JF) and whose Ly$\alpha$ lines penetrate $u$ ($u^\ast$) or $g$-band filter (Fig.~\ref{fig1}).
Following sections describe the databases and selection procedures.

\subsection{HSC-SSP--CLAUDS--DUNES2 joint field}
\label{s21}

This work is based on the HCD-JF data from multiple legacy surveys: the HSC-SSP (Public Data Release 2, PDR2; \citealt{Aihara2019}), CLAUDS \citep{Sawicki2019}, and DUNES$^2$ (Egami et al., in preparation).
The HCD-JF data platform provides a science-ready catalogue and coadd data, which have been jointly processed through the dedicated pipeline ({\tt hscPipe version 6}; \citealt{Bosch2018}) over 13 deg$^2$ in an effective area, by adapting advanced procedures developed by \citet{Desprez2023}. 
The survey area is split on the database into $1.7\times1.7$ deg$^2$ areas, termed {\tt tracts}, and further divided into $12\times12$ arcmin$^2$ regions called {\tt patches} \citep{Aihara2019}. 
This paper solely overviews how the HCD-JF data are assembled because the catalogue paper will give complete details of the data reduction and validation (HSC joint collaboration, in preparation).

\begin{figure*}
\centering
\includegraphics[width=12cm]{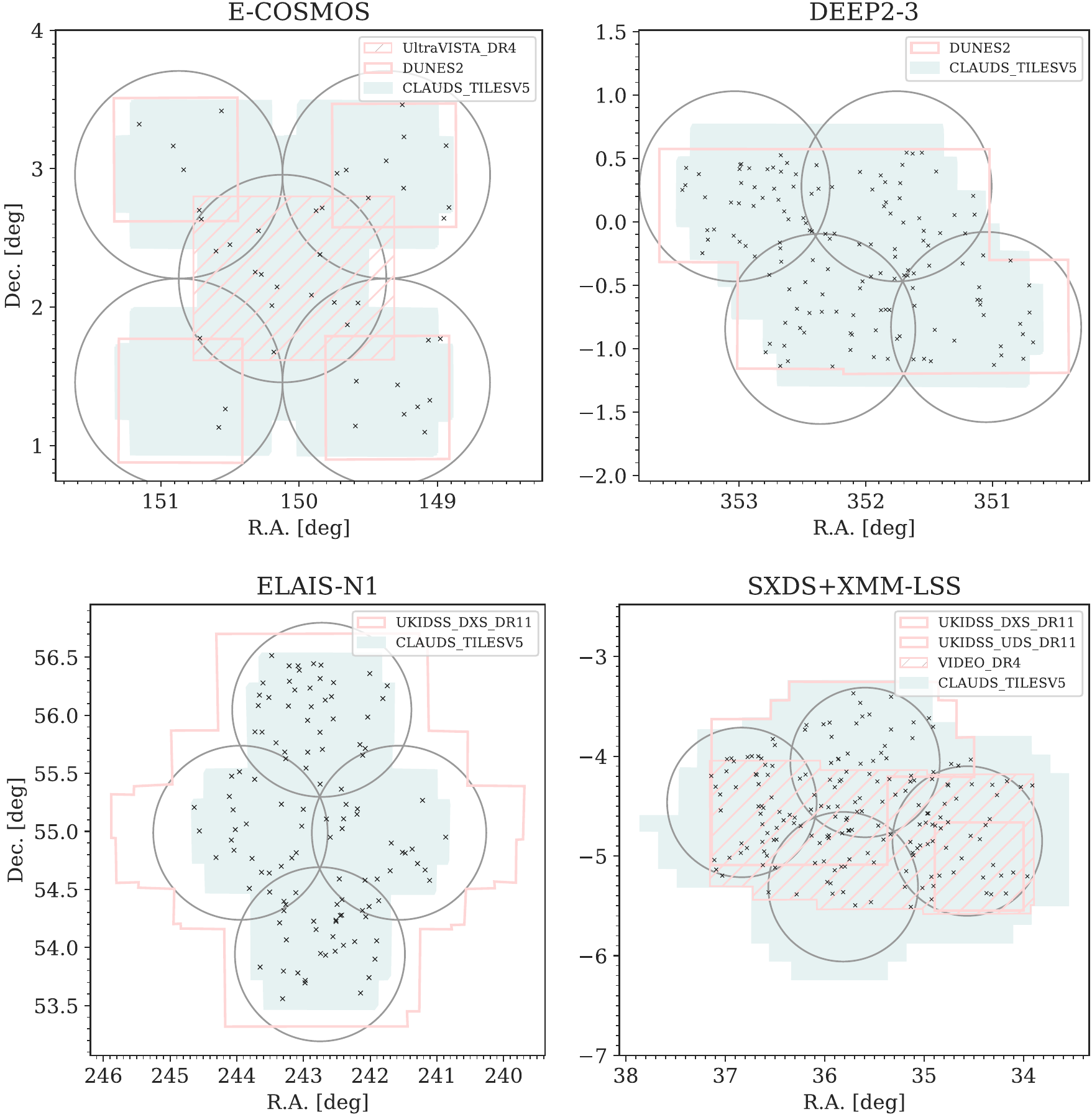}
\caption{
Survey fields of the HSC-SSP Deep layers (grey open circles), CLAUDS (cyan-filled regions), and a compilation of multiple near-infrared imaging surveys, including DUNES$^2$, UKIDSS, UltraVISTA, VIDEO (red open or red-hatched squares).
Black cross symbols depict our target quasars at $z=$ 1.9--3.0.
}
\label{fig2}
\end{figure*}

The HCD-JF is originally based on the HSC-SSP Deep and UltraDeep (DUD) layer over 36 deg$^2$ \citep{Aihara2018}, which targets four fields: E-COSMOS, DEEP2-3, ELAIS-N1, and SXDS+XMM-LSS (Fig.~\ref{fig2}).
Imaging depths in the $grizy$ bands of the HSC-SSP Deep layer are comparable to those of the Rubin/LSST \citep{Ivezic2019}.
Thus, following collaborating surveys provide complementary data in other wavebands.
In particular, CLAUDS and DUNES$^2$ programmes play imperative roles in extracting Ly$\alpha$ images and constraining photometric redshifts in this work.
The CLAUDS $u$ ($u^\ast$) -band data from the MegaCam imager on the Canada–France–Hawaii 3.6m telescope reach a median $5\sigma$ depth of 27.1 mag in 2 arcsec aperture diameter over 18.6 deg$^2$ out of the HSC-SSP DUD layer, which is by $\sim1$ mag deeper than that of Rubin/LSST \citep{Ivezic2019}.
Here, we adopt the $u^\ast$-band data in the SXDS+XMM-LSS field and the $u$-band data in the other fields.
Near-infrared data are delivered by multiple surveys: DUNES$^2$ (Egami et al., in preparation), UKIDSS Deep Extragalactic Survey in the ELAIS-N and XMM-LSS fields \citep{Lawrence2007}, UltraVISTA near-infrared imaging survey in the COSMOS field \citep{McCracken2012}, and VISTA Deep Extragalactic Observations (VIDEO) survey in the XMM-LSS field \citep{Jarvis2013}.
The combined HCD-JF data cover 13 deg$^2$ with at least $ugrizyJK$ bands by aforementioned surveys. 
Table~\ref{tab1} lists the dataset and typical imaging depths.

\begin{table}
\caption{
Data summary.
Area coverage is shown in Fig.~\ref{fig2}.
}
\label{tab1}
\begin{tabular}{cccc} 
\hline
Filter   & $\lambda$ ($\mu$m) & Depth$^a$ (mag) & Field$^b$\\
\hline
\multicolumn{4}{l}{CLAUDS \citep{Sawicki2019}}\\
$u$      & 0.37 & 27.1 & ECOS, ELN1, DE23\\
$u^\ast$ & 0.38 & 27.1 & ECOS, XMML\\
\hline
\multicolumn{4}{l}{HSC-SSP PDR2 \citep{Aihara2019}}\\
$g$      & 0.48 & 27.3 & ECOS, ELN1, DE23, XMML\\
$r$      & 0.62 & 26.9 & ECOS, ELN1, DE23, XMML\\
$i$      & 0.77 & 26.7 & ECOS, ELN1, DE23, XMML\\
$z$      & 0.89 & 26.3 & ECOS, ELN1, DE23, XMML\\
$y$      & 0.98 & 25.3 & ECOS, ELN1, DE23, XMML\\
\hline
\multicolumn{4}{l}{DUNES$^2$ (Egami et al. in preparation)}\\
$J$      & 1.25 & 23.2 & ECOS, DE23\\
$H$      & 1.63 & 23.2 & ECOS, ELN1\\
$K$      & 2.20 & 23.0 & ECOS, DE23\\
\hline
\multicolumn{4}{l}{UKIDSS DR11 \citep{Lawrence2007}}\\
$J$      & 1.25 & 22.3/24.8$^c$ & ECOS, ELN1\\
$H$      & 1.63 & 21.8/23.8$^c$ & XMML\\
$K$      & 2.20 & 20.8/22.8$^c$ & ECOS, ELN1\\
\hline
\multicolumn{4}{l}{VIDEO DR4 \citep{Jarvis2013}}\\
$Y$      & 1.02 & 24.6 & XMML\\
$J$      & 1.25 & 24.5 & XMML\\
$H$      & 1.65 & 24.0 & XMML\\
$K_s$    & 2.15 & 23.5 & XMML\\
\hline
\multicolumn{4}{l}{UltraVISTA DR4 \citep{McCracken2012}}\\
$Y$      & 1.02 & 24.7/25.8$^d$ & ECOS\\
$J$      & 1.25 & 24.5/25.6$^d$ & ECOS\\
$H$      & 1.65 & 24.1/25.2$^d$ & ECOS\\
$K_s$    & 2.15 & 24.5/24.9$^d$ & ECOS\\
\hline
\multicolumn{4}{l}{$^a$ Typical $5\sigma$ limiting magnitude in 2" apertures}\\
\multicolumn{4}{l}{$^b$ ECOS: E-COSMOS, ELN1: ELAIS-N1, DE23: DEEP2-3,}\\
\multicolumn{4}{l}{~~~XMML: XMM-LSS}\\
\multicolumn{4}{l}{$^c$ Depths in DXS/UDS layers}\\
\multicolumn{4}{l}{$^d$ Depths in Ultra-deep/Deep layers}\\
\end{tabular}
\end{table}

\subsection{Quasar samples}
\label{s22}

The sample selection basically follows our previous work \citep{Shimakawa2022}, apart from the fact that this work additionally employs quasars at lower redshift ($z=$ 1.9--2.23) for $u,u^\ast$-band selections (for definition, see Table~\ref{tab2} and Section~\ref{s3}).
First, we select 119,219 quasars at the Ly$\alpha$ redshift ({\tt Z\_LYA}) = 1.9--3.0 in the DR16Q catalogue whose Ly$\alpha$ emissions fall into $u$ ($u^\ast$) or $g$-band (Fig.~\ref{fig1}). 
Here, we do not adopt quasars at $z=$ 3.0--3.5 even though $g$-band captures their Ly$\alpha$ lines, because of larger uncertainties in the Ly$\alpha$ image extraction due to effects of foreground IGM absorption (see section~3 in \citealt{Shimakawa2022}).
The DR16 quasars are originally based on three datasets of SDSS-IV/eBOSS \citep{Myers2015}, the Wide-field Infrared Survey Explorer \citep{Wright2010,Lang2016}, and the Palomar Transient Factory \citep{Law2009,Rau2009} over 14,000 deg$^2$, with a magnitude limit of $g<22$ or $r<22$.
Therefore, quasar verification has been made based on their spectra from the eBOSS spectrographs \citep{Smee2013} by matching to quasar models through the BOSS {\tt spec1d} pipeline (\citealt{Bolton2012}; see sections~2 and 3 in \citealt{Lyke2020} for details).

\begin{table}
\caption{SDSS/eBOSS quasars adopted in this study.}
\label{tab2}
\begin{tabular}{lllllll} 
	\hline
	Band$^1$ & N & N$_\mathrm{nebula}^2$ & Redshift & Thr$^3$ & Lim$^4$ & FWHM$^5$\\
	\hline
	$ugr$       & 147 &  8 & 1.90--2.23 & $2.6$ & $4.0$ & 1.1"\\
	$u^\ast gr$ & 120 & 10 & 1.90--2.23 & $3.2$ & $5.4$ & 1.1"\\
	$gri$       & 216 &  6 & 2.34--3.00 & $4.7$ & $6.6$ & 1.0"\\
	\hline
\multicolumn{7}{l}{$^1$ Selection bands for Ly$\alpha$ and rest-frame UV images in \S\ref{s3}.}\\
\multicolumn{7}{l}{$^2$ Number of quasars with large nebulae used in \S\ref{s5}.}\\
\multicolumn{7}{l}{$^3$ $\mathrm{SB_{Ly\alpha}}$ thresholds ($2\sigma$) per binned pixel in $10^{-18}$ erg~s$^{-1}$cm$^{-2}$arcsec$^{-2}$}\\
\multicolumn{7}{l}{$^4$ $\mathrm{SB_{Ly\alpha}}$ limits ($2\sigma$) in 1 arcsec$^2$ aperture in $10^{-18}$ erg~s$^{-1}$cm$^{-2}$arcsec$^{-2}$}\\
\multicolumn{7}{l}{$^5$ Matched seeing FWHM in arcsec}\\
\end{tabular}
\end{table}

Next, we cross-match the quasar samples with the HCD-JF sources within a radius of 1 arcsec.
Here, we exclude samples that do not match the seeing criteria in either of the selection bands (Table~\ref{tab2}), and the seeing sizes of the broad-band images used in each selection have been matched to these thresholds.
One should note that selected quasars are not affected by bad pixels and not located at the edge of the survey fields in the $ugrizy$-bands by applying the following criteria in the {\tt SQL} query:
\begin{description}
\item[--]{\tt isprimary=True},
\item[--]{\tt inputcount\_flag\_noinputs=False},
\item[--]{\tt pixelflags\_edge=False},\
\item[--]{\tt pixelflags\_bad=False},
\end{description}
Refer to \citet{Coupon2018,Bosch2018,Aihara2019} for details on these catalogue flags. 
Consequently, we collect a total of 483 quasars at $z=$ 1.9--3.0 in the HCD-JF, where 147, 120, and 216 sources are selected for the $ugr$, $u^\ast gr$, and $gri$ selections, respectively (Table~\ref{tab2}).
Quasar's identification numbers and sky coordinates from the HSC-SSP data release ({\tt s18a\_dud\_u2k}) and Ly$\alpha$ redshifts ({\tt Z\_PIPE} and {\tt Z\_LYA} from \citealt{Lyke2020}) are summarised in Table~\ref{tab3}.

Following the target selection, we cut out coadd images ($200\times200$ pixel$^2$ with a pixel scale of 0.168 arcsec) for the selected quasars from the HCD-JF database.
Similar to the preprocessing in \citet{Shimakawa2022}, we match the seeing sizes of individual images to those in Table~\ref{tab2}, depending on the selection filters, and then carry out the spatial $2\times2$ binning to increase the signal-to-noise ratio (SNR).
Therefore, final coadd data used in this study are organised by the seeing-matched images of $100\times100$ pixel$^2$ with a pixel scale of 0.336 arcsec.
The image size corresponds to 290--265 proper kpc (pkpc) at $z=$ 1.9--3.0.
Galactic extinctions are also corrected based on \citet{Schlegel1998}.

\begin{table*}
\caption{
Our quasar targets and their properties used in this work. 
The full source catalogue is available as online material in CSV format.
}
\label{tab3}
\begin{tabular}{cccccccccccc}
    \hline
    {\tt object\_id} & ra & dec & {\tt Z\_PIPE}$^1$ & {\tt Z\_LYA}$^1$ & {\tt M\_I}$^1$ & Colour & {\tt Area} & {\tt Area\_eff} & {\tt Area\_thr} & $d_\mathrm{max}$ & $\alpha$ \\
    & & & & & & & (arcsec$^2$) & (arcsec$^2$) & (arcsec$^2$) & (pkpc) &\\
    \hline
    37484563299064740 & 34.93038 & -5.43555 & 2.788 & 2.856 & -26.08 & $gri$       & 20.9 & 15.8 & 11.3 & --- & ---\\
    37484563299081621 & 34.91202 & -5.30154 & 2.226 & 2.219 & -26.37 & $u^\ast gr$ & 83.5 & 75.8 & 37.6 & 135 & 0.73\\
    37484567594035565 & 34.93243 & -5.20900 & 2.027 & 2.025 & -23.40 & $ugr$       &  9.5 &  6.2 &  2.3 & --- & ---\\
    \multicolumn{12}{c}{... (483 rows in total)}\\
    \hline
    \multicolumn{12}{l}{$^1$ Excerpts from \citet{Lyke2020}}\\
\end{tabular}
\end{table*}


\section{Broad-band colour selection}
\label{s3}

We search for extended Ly$\alpha$ emission around the selected 483 quasars using the broad-band selection, as demonstrated by \citet{Shimakawa2022}.
The selection procedure is quite similar to the traditional narrow-band $+$ broad-band technique (e.g., \citealt{Bunker1995}); however, we instead utilise three broad-band images as follows:
\begin{enumerate}
  \item Preparing Ly$\alpha$-band and two rest-frame UV images (Table~\ref{tab2}), 
  \item Extrapolating continuum at Ly$\alpha$-band from two broad-bands,
  \item Extracting Ly$\alpha$ images by subtracting the obtained continuum image from the observed Ly$\alpha$-band image, 
  \item Object masking of irrelevant sources based on photo-$z$.
\end{enumerate}
Here, the Ly$\alpha$-band means the filter band that contains Ly$\alpha$ emission (Fig.~\ref{fig1}), corresponding to the $u$ ($u^\ast$) -band for quasars at $z=$ 1.90--2.23 or $g$-band for quasars at $z=$ 2.34--3.00.
We also adopt $g,r$-bands and $r,i$-bands, respectively, to construct their continuum images at the Ly$\alpha$ wavelength (Table~\ref{tab2}).
Notably, this study ignores potential contributions from rest-UV helium and metal lines to the broad-band images, which could lead to over- or under-estimates of Ly$\alpha$ emission at some level depending on the source redshift.
However, these emission lines are generally an order of magnitude fainter than Ly$\alpha$ line (see, e.g., \citealt{Cai2017,Guo2020}) and are expected to have little effect on our selection.
Furthermore, given that there may be foreground IGM absorption effects, we need to bear such possible systematic errors in mind when we discuss the measurement accuracy. 
\citet{Shimakawa2022} have noted that increasing uncertainties of the extrapolation of the continuum band at $z>3$ because strong IGM absorption at $\lesssim1020$ \AA\ penetrates the $g$-band. 
Therefore, we remove quasars at $z>3$ from the sample (Table~\ref{tab2}).

\begin{figure*}
\centering
\includegraphics[width=17cm]{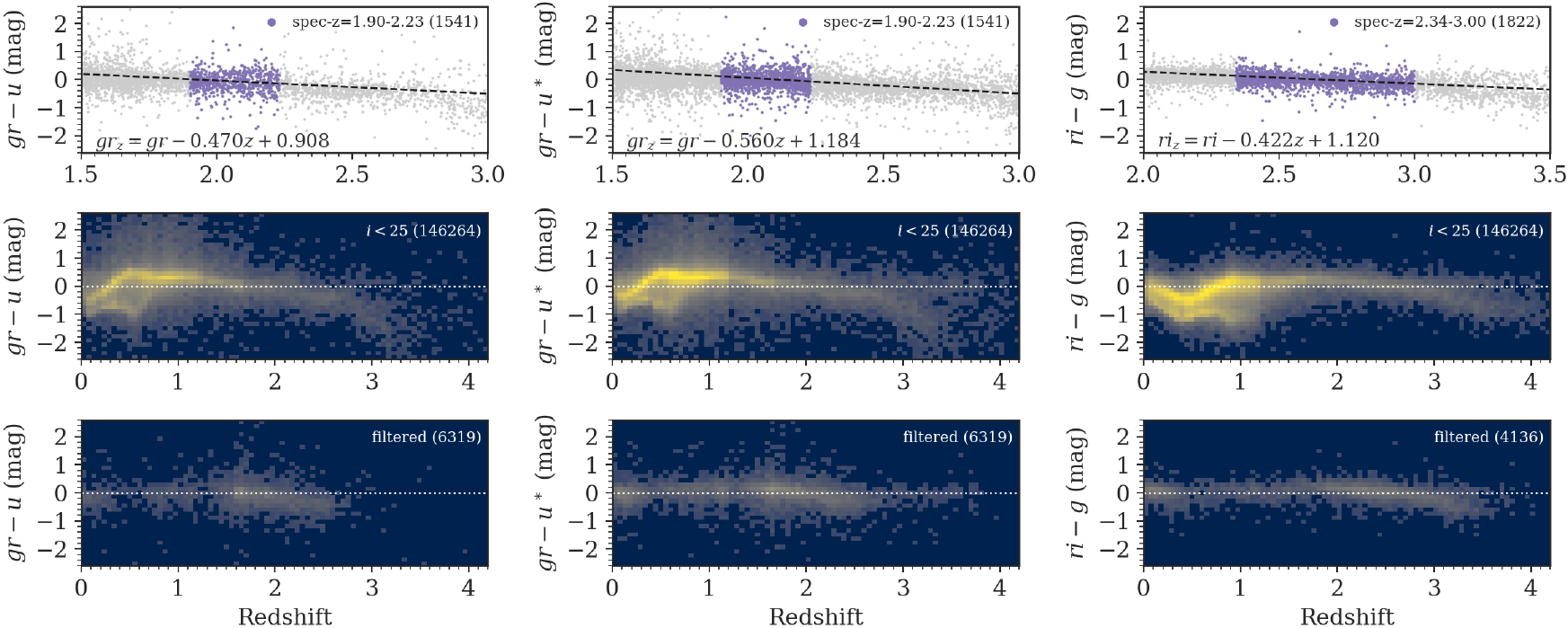}
\caption{
Redshift distributions of the colour terms in the three broad-band selections, from left to right: $gr-u$, $gr-u^\ast$, and $ri-g$.
In the first row, purple and grey dots indicate spec-$z$ sources at the target redshift in each selection (Table~\ref{tab2}) and sources in the foreground and background, respectively.
The second row shows colour-term distributions for all spec-$z$ sources at $z=$ 0--4 ($N=$ 146,264).
The third row is identical to the second, but shows the filtered spec-$z$ samples based on their photometric redshifts.
We here exclude photo-$z$ sources that do not fall into the target redshift ranges (Table~\ref{tab2}) within $|\delta z|<0.1$ (see also text).
}
\label{fig3}
\end{figure*}

The following explains how we establish the continuum image and mask potential contaminants, which are technically the most important in the selection process.
We must remove continuum components of host galaxies and satellites and mask irrelevant sources to extract extended Ly$\alpha$ emission associated with quasars from the broad-band image containing Ly$\alpha$ emission (Ly$\alpha$-band).
Therefore, as shown in Fig.~\ref{fig3}, we empirically extrapolate the continuum image compatible with the Ly$\alpha$-band from the two rest-frame UV images in each selection (Table~\ref{tab2}, see also \citealt{Shimakawa2022}).
Specifically, we derive the best-fit conversion factor from the rest-frame UV band magnitudes ($gr$ or $ri$) to Ly$\alpha$-band magnitudes ($u$ or $g$) for known spec-$z$ sources.
\begin{eqnarray}
    gr &=& 1.763\ g - 0.754\ r \qquad (\mathrm{for}\ u\mbox{-}\mathrm{band}),\\
    gr &=& 1.633\ g - 0.618\ r \qquad (\mathrm{for}\ u^\ast\mbox{-}\mathrm{band}),\\
    ri &=& 1.350\ r - 0.338\ i \qquad (\mathrm{for}\ g\mbox{-}\mathrm{band}).
\label{eq1}
\end{eqnarray}
For each regression, we collect 1,541--1,822 spec-$z$ samples from the HCD-JF database (see the numbers in Fig.~\ref{fig3}): 3D-HST \citep{Brammer2012,Momcheva2016}, DEIMOS 10k sample \citep{Hasinger2018}, GAMA DR3 \citep{Baldry2018}, PRIMUS \citep{Coil2011,Cool2013}, SDSS DR15 \citep{Aguado2019}, and VVDS \citep{LeFevre2013}.
Here, we remove the SDSS/eBOSS quasars from the spec-$z$ sample.
The best-fit relations are obtained using a non-linear optimisation and curve-fitting tool for Python, {\tt lmfit} (version 1.0.3; \citealt{Newville2014}). 
Although Ly$\alpha$ emission of the spec-$z$ samples, especially Ly$\alpha$ emitters, might contribute to their Ly$\alpha$-band photometry, we assume that, overall, their Ly$\alpha$ emissions do not significantly affect the fitting.
The best-fit relations well reproduce the continuum fluxes in nearby redshift ranges, where there are no Ly$\alpha$ contributions to the Ly$\alpha$-bands (Fig.~\ref{fig3}).
Additionally, we correct a small redshift ($z$) dependence of the colour-term as depicted in Fig.~\ref{fig3},
\begin{eqnarray}
    gr_z &=& gr - 0.470\ z - 0.908 \qquad (\mathrm{for}\ u\mbox{-}\mathrm{band}),\\
    gr_z &=& gr - 0.560\ z - 1.184 \qquad (\mathrm{for}\ u^\ast\mbox{-}\mathrm{band}),\\
    ri_z &=& ri - 0.422\ z + 1.120 \qquad (\mathrm{for}\ g\mbox{-}\mathrm{band}).
\label{eq2}
\end{eqnarray}
We used $gr_z$ and $ri_z$-band images to subtract the continua from the $u$ ($u^\ast$) and $g$-band images.

\begin{figure}
\centering
\includegraphics[width=8cm]{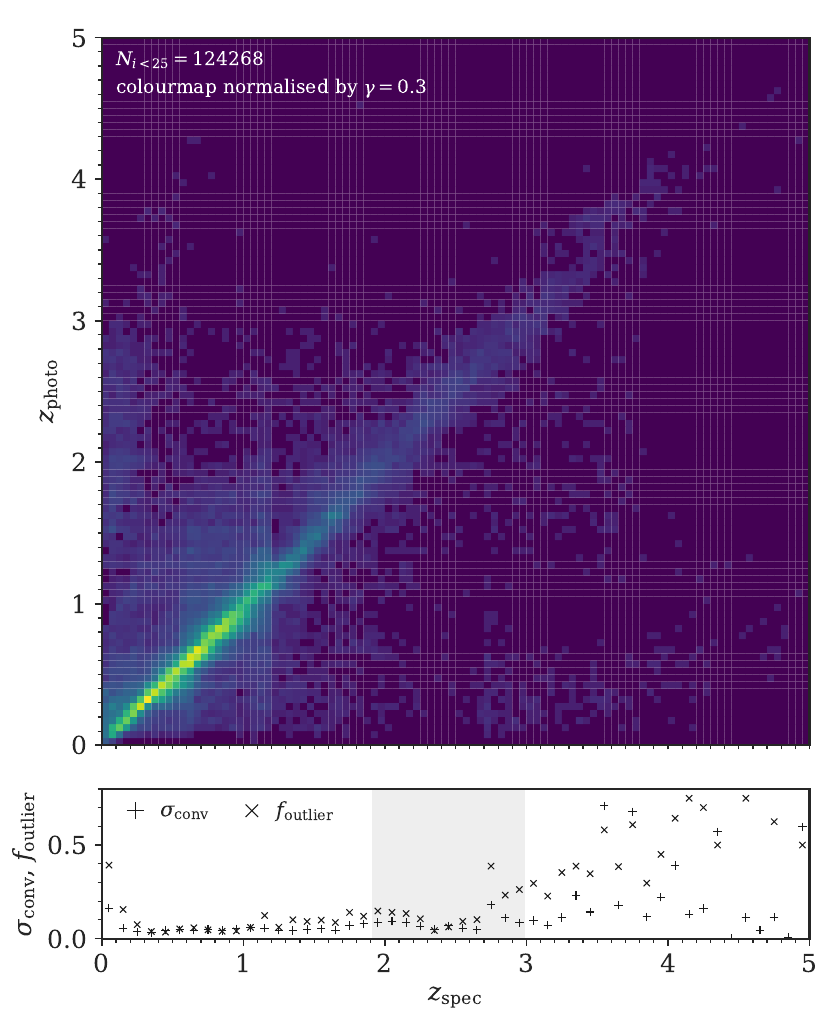}
\caption{
(Top) Photometric redshifts versus spectroscopic redshifts for 124,268 spec-$z$ sources ($i<25$ mag), with additional selection thresholds ({\tt stellar\_mass}$>10^{7.5}$ and {\tt photoz\_risk\_best}$<0.2$) consistent with those in \citet{Tanaka2018}.
(Bottom) Dispersion ($\sigma_\mathrm{conv}$) and outlier rates ($f_\mathrm{outlier}$) as a function of spectroscopic redshifts (see text for details).
The shaded area depicts the target redshift range ($z=$ 1.9--3.0). 
}
\label{fig4}
\end{figure}

Fig.~\ref{fig3} illustrates the colour-term distributions of a total of 146,264 spec-$z$ sources at $z=$ 0--4 in three different colour selections ($gr-u$, $gr-u^\ast$, and $ri-g$).
Importantly, we must filter projected neighbours in the foreground and background, especially at $z<2$, which may cause pseud Ly$\alpha$ features. 
Otherwise, all positive colour values in the figure will present as Ly$\alpha$ emission in the residual images.
Therefore, we employ a photometric redshift catalogue based on the SED-fitting code {\tt Mizuki} \citep{Tanaka2015,Tanaka2018} to mask foreground and background sources within individual images.
With the aid of $u$ and NIR photometry, we can well constrain photometric redshift up to $z\sim3$ compared to the original photo-$z$ estimation only with the HSC $grizy$ filters (see also the validation by \citealt{Desprez2023}).
Median dispersion ($\sigma_\mathrm{conv}$) and outlier rate ($f_\mathrm{outlier}$) for $i<25$ sources at $z=$ 1.9--3.0 are estimated to be 0.08 and 0.13, respectively, according to the available spec-$z$ samples from the HCD-JF data (Fig.~\ref{fig4}).
Obtained dispersion and outlier rate are approximately half and one-third of those from only the HSC filters, respectively (see table~3 in \citealt{Tanaka2018}). 
The photo-$z$ dispersion and outlier rate are defined as follows \citep[eq.~10-11]{Tanaka2018}:
\begin{eqnarray}
    \sigma_\mathrm{conv} &=& 1.48~\mathrm{MAD}(\delta z),\\
    f_\mathrm{outlier} &=& \frac{N(|\delta z|>0.15)}{N_\mathrm{total}},
\label{eq3}
\end{eqnarray}
where $\delta z$ is a systematic offset from spectroscopic redshift defined by $\delta z=(z_\mathrm{photo}-z_\mathrm{spec})/(1+z_\mathrm{spec})$, and MAD is the median absolute deviation.
We also confirm that photo-$z$ distributions are broadly consistent across the four survey fields, and there is no peculiar trend depending on the available filter sets.
Taking account of the photo-$z$ errors, we mask $i$-band magnitude limited sources ($i<25$) at foreground and background redshifts with $|\delta z|>0.1$ from the quasar redshift ({\tt Z\_PIPE} in \citealt{Lyke2020}, i.e., $|(z_\mathrm{photo}-z_\mathrm{PIPE})/(1+z_\mathrm{PIPE})|>0.1$).
For reference, the colour-term distribution after filtering potential contaminants out of the selection redshifts (Table~\ref{tab2}) can be found in the bottom panel of Fig.~\ref{fig3}.
Specifically, over the whole targets, approximately 95\% of $i$-band magnitude limited sources ($i<25$) are masked as irrelevant foreground and background sources based on photometric redshifts. 
We confirm that our object masking is sufficiently functioning particularly for those at $z\lesssim1.6$, the major contaminants in the $ugr$ and $u^\ast gr$ selections as seen in the middle row of Fig.~\ref{fig3}, where 99\% of such foreground sources are successfully rejected according to the spec-$z$ samples.

\begin{figure*}
\centering
\includegraphics[width=17.5cm]{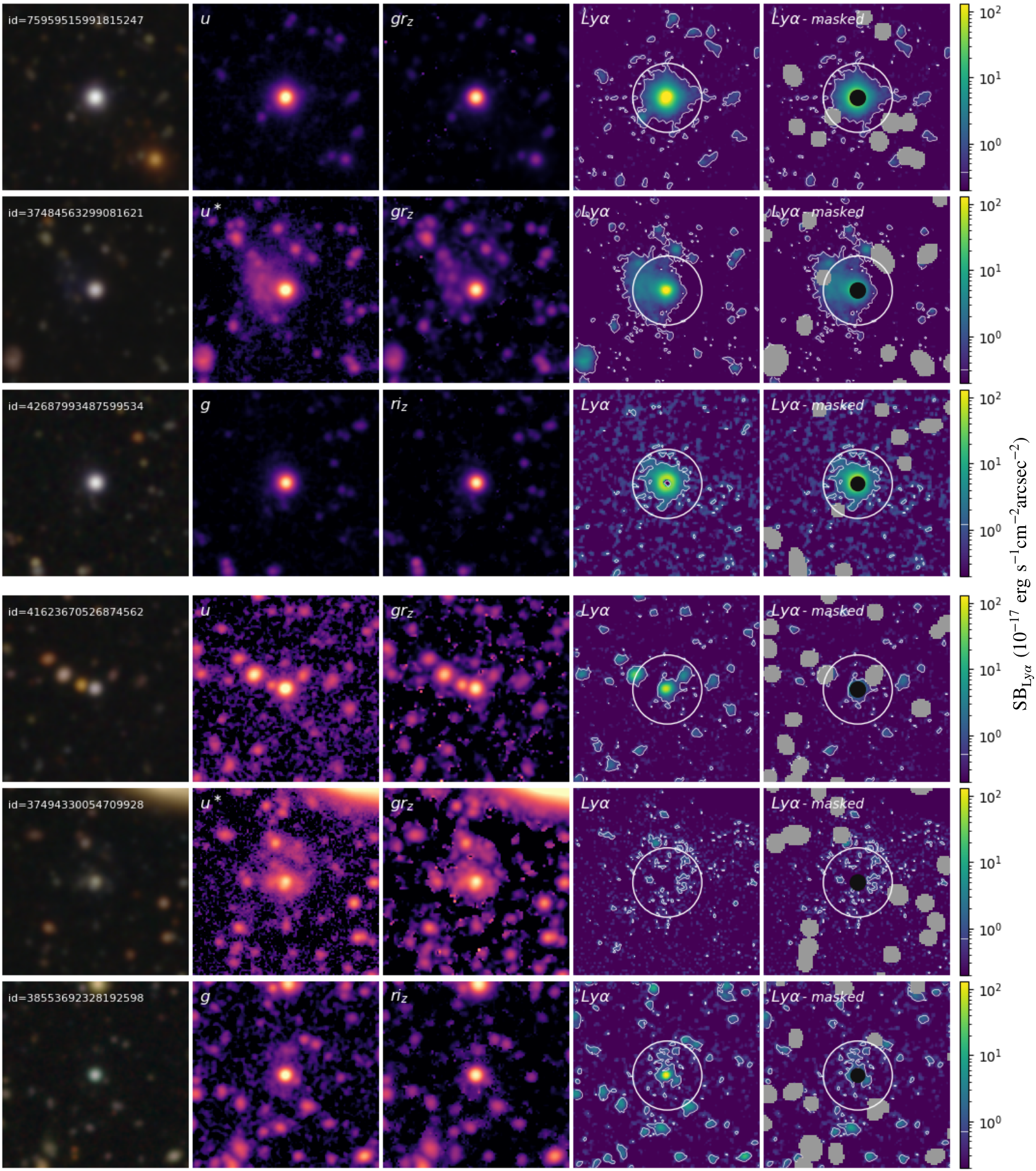}
\caption{
From left to right, RGB colour cutouts, broad-band coadd images with Ly$\alpha$ contributions (Ly$\alpha$-bands: $u$, $u^\ast$, $g$), extrapolated broad-band images ($gr_z$, $gr_z$, $ri_z$), and Ly$\alpha$ maps ($gr_z-u$, $gr_z-u^\ast$, $ri_z-g$) without and with masks.
The colour maps in the fourth and fifth columns are scaled by Ly$\alpha$ surface brightness (SB$_\mathrm{Ly\alpha}$) as depicted by the right colour bars (in the unit of $10^{-17}$ erg~s$^{-1}$cm$^{-2}$arcsec$^{-2}$).
The top three rows, from top to bottom, show quasar nebulae with the largest {\tt Area\_thr} in the $ugr$, $u^\ast gr$, and $gri$ selections ({\tt object\_id} = 75959515991815247, 37484563299081621, 42687993487599534 in Table~\ref{tab3}).
On the other hand, the bottom three show examples of non quasar nebulae that do not meet the selection criteria ({\tt object\_id} = 41623670526874562, 37494330054709928, 38553692328192598). 
Here, we pick up a quasar sample in each selection that seems to have nebula from its $u$-band image alone, but do not detect extended Ly$\alpha$ emission at a significant level.
The images span 34 arcsec ($\sim260$ pkpc), and the white contour depicts a $2\sigma$ excess above the background deviation ({\tt bg\_std}). 
White open circles mark the maximum radius used to measure {\tt Area}, {\tt Area\_eff}, and {\tt Area\_thr} \citep[fig.~7]{Shimakawa2022}.
Grey regions and the black centre circle in the right panel are object masks applied to remove foreground or background contaminants and quasar itself, respectively.
Similar diagrams for all 483 quasar samples are available as online material.
}
\label{fig5}
\end{figure*}

Fig.~\ref{fig5} illustrates image examples, starting from the left, the RGB colour, Ly$\alpha$-band, extrapolated continuum, Ly$\alpha$ map, and Ly$\alpha$ map with object masking for the targets in three different colour selections. 
The processed images for all 483 quasars (see Section~\ref{s41}) are available as a supplemental material, allowing readers to visually check spatial distributions of extracted Ly$\alpha$ emission and possible contaminants from projected neighbours for individual quasars.
The mask areas to irrelevant neighbours at $|\delta z|>0.1$ are determined by $2.5\times$ PSF-convolved major- and minor-axis based on the second moments of the object intensity, termed adaptive moments in the $i$-band ({\tt i\_sdssshape\_shape} in the HSC-SSP database; see also \citealt{Bernstein2002}).
In addition, we mask bright objects with $i<18$ mag and point sources regardless of their photometric redshifts ({\tt i\_psfflux\_mag$-$i\_cmodel\_mag$<0.2$}; see \citealt{Strauss2002,Baldry2010}).
Moreover, quasars are masked within 2.5 arcsec diameter ($2.5\times$ seeing FWHM or $r\lesssim10$ pkpc) for relatively fair comparisons of Ly$\alpha$ properties between optically bright and faint quasars.
Recent IFU studies have attempted to subtract the quasar components by applying more sophisticated PSF subtraction technique \citep{Cantalupo2019}. 
However, in the broad-band selection, we encounter cases, such as those shown in the third and fifth rows of Fig.~\ref{fig5}, where the PSF modelling proves challenging. 
Therefore, we opt for a simple approach by masking the centre at this time. 
Such a limitation remains an issue to be addressed in future work.

Finally, we should describe a strong limitation of the broad-band selection. 
Although this method is basically the same as the narrow-band technique, it is only effective for sources with high equivalent widths (EW$_\mathrm{Ly\alpha}$) because it is less sensitive to the emission line contrast due to the wider filter bandwidths. 
The specific threshold varies depending on a combination of the uncertainty of the colour term correction (Eq.~1--6) and filter bandwidth ($\sim700$~\AA\ or $\sim1400$~\AA). 
In this study, it is estimated to be approximately 100--200\AA\ in the rest-frame with a $2\sigma$ excess. 
It suggests that our selection method cannot probe Ly$\alpha$ emissions for most of Ly$\alpha$ emitters with EW$_\mathrm{Ly\alpha}\lesssim100$~\AA\ at a significance level (see e.g., \citealt{Gronwall2007,Cassata2015,Hashimoto2017,Kerutt2022}), but can detect those from active galactic nuclei, quasars, or bright Ly$\alpha$ nebulae in the outskirts of galaxies, where EW$_\mathrm{Ly\alpha}$ can be detectably high (e.g., \citealt{Rakshit2020,Liu2022}).


\section{Quasar nebulae}
\label{s4}

\subsection{Spatial extent}
\label{s41}

\begin{figure}
\centering
\includegraphics[width=8cm]{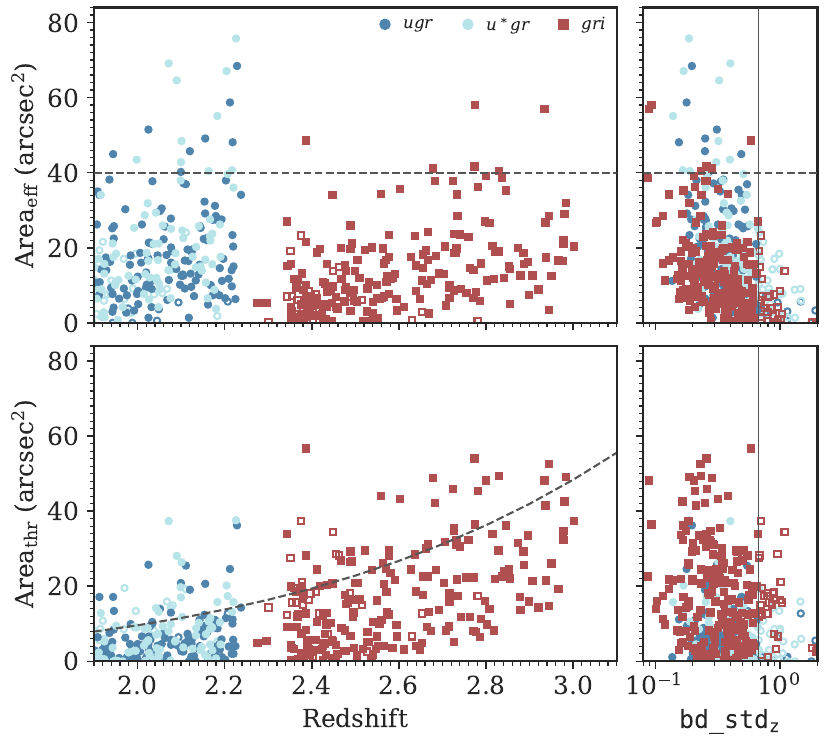}
\caption{
Ly$\alpha$ areas of 483 quasars plotted against (left) redshift and (right) the standard deviation of redshift dimming-corrected background flux ({\tt bg\_std$_z$}) in the unit of $10^{-17}$ erg~s$^{-1}$cm$^{-2}$arcsec$^{-2}$. 
(Top) Effective Ly$\alpha$ areas ({\tt Area\_eff}), accounting for object masking of projected neighbours and the quasar itself.
(Bottom) Effective Ly$\alpha$ areas ({\tt Area\_thr}) above a redshift-dimming corrected SB$_\mathrm{Ly\alpha}>1\times10^{-17}$ erg~s$^{-1}$cm$^{-2}$arcsec$^{-2}$ at $z=2.3$.
In the right panels, solid vertical lines depict the selection criteria of {\tt bg\_std$_z$} $<0.67\times10^{-17}$ erg~s$^{-1}$cm$^{-2}$arcsec$^{-2}$, samples above which (open symbols) are not considered in the main analyses. 
The dashed horizontal lines in the top panels mark the selection criteria for morphological analysis (Section~\ref{s51}).
The dashed curve in the bottom left panel shows the $\sim$16th-percentile borderline as a function of redshift, used in Section~\ref{s52}.
}
\label{fig6}
\end{figure}

This section examines Ly$\alpha$ properties of the quasars probed by the broad-band selection.
Following \citet{Shimakawa2022}, this study defines Ly$\alpha$ areas ({\tt Area}) and effective Ly$\alpha$ areas with the object masking ({\tt Area\_eff}) above the two sigma in the standard deviation of the background ({\tt bg\_std}) within a diameter of 12 arcsec.
The standard deviations are based on pixel counts taken from backgrounds outside the radius of more than 12.5 arcsec ($\gtrsim100$ pkpc).
More details are illustrated in \citealt[fig.~7]{Shimakawa2022}.
Similarly, we define a redshift dimming-corrected Ly$\alpha$ area ({\tt Area\_thr}) within a diameter of 12 arcsec above the certain Ly$\alpha$ surface brightness, SB$_\mathrm{Ly\alpha}>1\times10^{-17}$ erg~s$^{-1}$cm$^{-2}$arcsec$^{-2}$ at $z=2.3$, i.e., surface brightness scaled by $(1+z)^4/(1+2.3)^4$, because this study manages quasars in different redshift ranges ($z=$ 1.9--3.0) based on the three colour selections.
These obtained parameters for all quasars are listed in Table~\ref{tab3}.
Ly$\alpha$ surface brightness ($\mathrm{SB_{Ly\alpha}}$) in each pixel is measured from the Ly$\alpha$ image as follows:
\begin{equation}
\mathrm{SB_{Ly\alpha}}=\Delta_\mathrm{FWHM}\cdot w(z)\cdot f_\mathrm{Ly\alpha}/A ~~~\mathrm{(erg~s^{-1}cm^{-2}arcsec^{-2})},
\label{eq9}
\end{equation}
where $\Delta_\mathrm{FWHM}$ is FWHM of the Ly$\alpha$ band ($u$: 650~\AA, $u^\ast$: 700~\AA, $g$: 1375~\AA), and $w(z)$ is a correction factor to compensate the filter transmittance depending on the Ly$\alpha$ redshift (Fig.~\ref{fig1}), which is normalised by the mean transmittance in each filter. 
We refer to quasar's Ly$\alpha$ redshits ({\tt Z\_LYA}) in \citet{Lyke2020} for the filter correction factor.
Furthermore, $f_\mathrm{Ly\alpha}$ is flux density in the Ly$\alpha$ band, and $A$ is a pixel area ($0.336\times0.336$ arcsec$^2$).
Example two-dimensional distributions of Ly$\alpha$ surface brightness around quasars are presented by the colour maps in Fig.~\ref{fig5}.
Notably, derived SB$_\mathrm{Ly\alpha}$ values, and, thus Ly$\alpha$ areas depend on various factors such as imaging depths ({\tt bg\_std}), contamination by projected neighbours in the foreground and background (Fig.~\ref{fig3}), and colour term effects ($\sim20$ per~cent in SB$_\mathrm{Ly\alpha}$; \citealt{Shimakawa2022}).
Therefore, as these individual properties are only provisional, the results in the following sections should be viewed in the context of such possible complex effects.
To ease the impact of poor quality data, this work does not use the samples with redshift-dimming corrected {\tt bg\_std}, {\tt bg\_std$_z$} $>0.67\times10^{-17}$ erg~s$^{-1}$cm$^{-2}$arcsec$^{-2}$, amounting to 74 sources as indicated by open symbols in Fig.~\ref{fig6}.

\begin{figure}
\centering
\includegraphics[width=8cm]{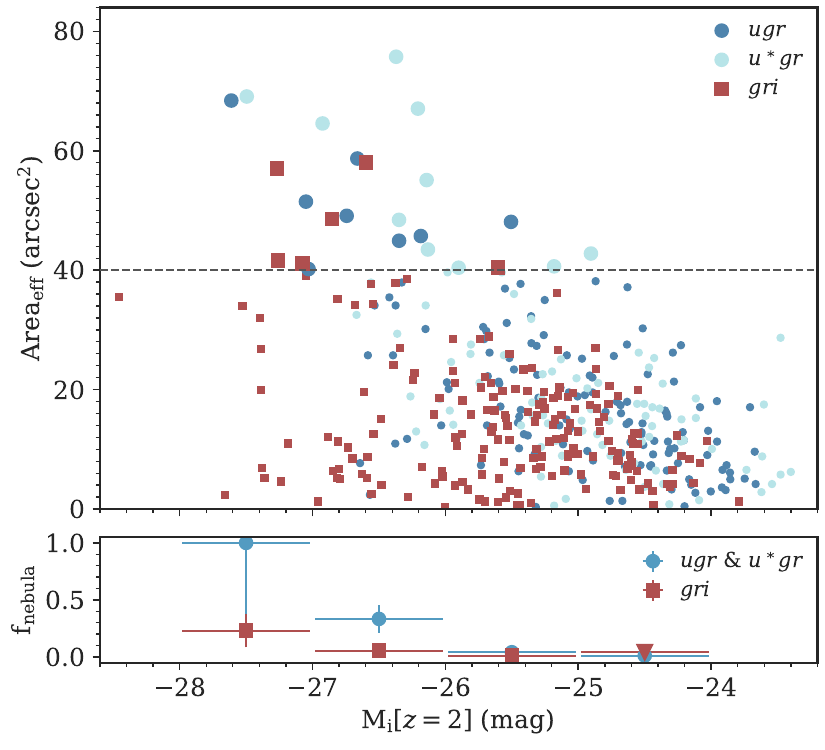}
\caption{
(Top) Effective Ly$\alpha$ areas ({\tt Area\_eff}) versus absolute $i$-band magnitudes normalised at $z=2$, $\mathrm{M_i}[z=2]$ taken from \citet{Lyke2020}.
Although symbols are same as in Fig.~\ref{fig6}, the selected quasar nebulae with effective Ly$\alpha$ areas $>40$ arcsec$^2$ are emphasised by the larger markers. 
(Bottom) Fractions of quasar nebulae in the $ugr$ and $u^\ast gr$ selections (cyans) and the $gri$ selection (reds).
The error bars show Poisson errors.
}
\label{fig7}
\end{figure}

Fig.~\ref{fig6} represents obtained effective Ly$\alpha$ areas ({\tt Area\_eff}) and redshift dimming-corrected Ly$\alpha$ areas ({\tt Area\_thr}) as a function of quasar's redshifts \citep[{\tt Z\_PIPE}]{Lyke2020}.
For the remaining 409 targets, we select quasars with effective Ly$\alpha$ areas, {\tt Area\_eff} $>40$ arcsec$^2$, to investigate spatial properties of Ly$\alpha$ nebulocities associated with quasars in the following sections.
The effective area threshold approximately corresponds to the circularised radius of 30 pkpc taking into account of the central mask, which is sufficiently away from the centre at a radius of $\lesssim20$ pkpc, where quasar's contributions would be non-negligible (see Section~\ref{s42}).
Hereafter, we define these selected quasars as `quasar nebulae', amounting to 8, 10, and 6 quasars in the $ugr$, $u^\ast gr$, and $gri$ selections, respectively (Table~\ref{tab2}).
The two times narrower filter width of $u$ ($u^\ast$) -band enables deeper Ly$\alpha$ imaging in the $ugr$ ($u^\ast gr$) selection at $z=$ 1.90--2.23 relative to the $gri$ selection at $z=$ 2.34--3.00.
Such a discrepancy of Ly$\alpha$ depths leads to higher fractions of extended Ly$\alpha$ nebulae in effective Ly$\alpha$ areas in the $u$ ($u^\ast$) -band selection as seen in the top panel of Fig.~\ref{fig6} (and Fig.~\ref{fig7}).
The two $u$-band selections ($ugr$ and $u^\ast gr$) have a similar wavelength coverage and imaging depth, showing the same trend in the Ly$\alpha$ areas given source redshifts (Fig.~\ref{fig6}).
Henceforth, quasar nebulae selected from $ugr$ and $u^\ast gr$ colours are treated as the same group ($u$-select) against the $g$-selected quasar nebulae at $z=$ 2.34--3.00.
In addition, we confirm a systematic increase of the redshift dimming-corrected Ly$\alpha$ areas ({\tt Area\_thr}) from $z=1.9$ to $z=3.0$ (see also Section~\ref{s42}), which is consistent with the increasing Ly$\alpha$ surface brightness of quasar nebulae from $z=2$ to $z=3$ in previous IFU surveys \citep{ArrigoniBattaia2019,Cai2019}.

Fig.~\ref{fig7} compares effective Ly$\alpha$ areas ({\tt Area\_eff}) with the absolute $i$-band magnitudes normalised at $z=2$ (M$\mathrm{_i[z=2]}$ from \citealt{Lyke2020}) for the quasar targets, in order to investigate selection effects of quasar nebulae over the whole sample. 
This comparison is also informative for understanding differences from quasar nebulae studied by previous IFU surveys \citep{Borisova2016,ArrigoniBattaia2019,Cai2019}, particularly in terms of their magnitude limited target selections (M$\mathrm{_i[z=2]}\lesssim-27$). 
Fig.~\ref{fig7} indicates that the selected quasar nebulae tend to be associated with the brighter quasars within the sample, where three brightest quasars with M$\mathrm{_i[z=2]}<-27$ in the $ugr$ and $u^\ast gr$ selections are all selected as quasar nebulae with {\tt Area\_eff} $>40$ arcsec$^2$, although only five sources are selected among 324 quasars at M$\mathrm{_i[z=2]}>-26$ mag.
Such a selection trend is consistent with findings in earlier studies \citep{Mackenzie2021,Shimakawa2022}, which report that brighter quasars tend to have Ly$\alpha$ nebulae with higher surface brightness.

\subsection{Radial profile}
\label{s42}

In this section, we investigate typical Ly$\alpha$ radial profiles of 24 quasar nebulae and then check potential covering fractions of Ly$\alpha$ detection for a given radius using the Monte Carlo simulation.
Here, we discuss them by dividing the sample into 18 $u$-selected and 6 $g$-selected quasar nebulae (Table~\ref{tab2}), whose median redshifts are $z=2.14$ and $z=2.78$, respectively.

\begin{figure}
\centering
\includegraphics[width=8cm]{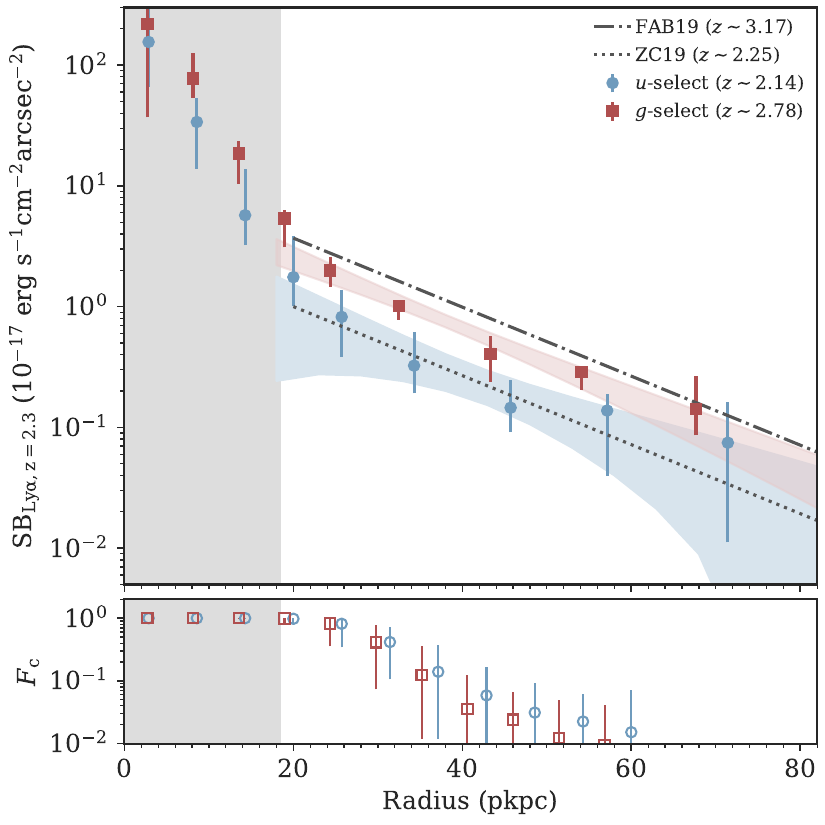}
\caption{
(Top) Radial profiles of Ly$\alpha$ surface brightness scaled by $(1+z)^4/(1+2.3)^4$.
Blue circles and red squares show the median radial profiles of $u$-selected and $g$-selected quasar nebulae, respectively, with error bars indicating the 68th-percentile deviations. 
The shaded regions show the 68th-percentile uncertainties in the exponential fit.
Black dotted and dash-dot lines represent the best-fit radial profiles of bright quasars at $z\sim2.25$ and $z\sim3.17$ from previous IFU surveys (ZC19: \citealt{Cai2019}, FAB19: \citealt{ArrigoniBattaia2019}).
The grey shaded area is masked in the fitting.
(Bottom) Median covering fractions of Ly$\alpha$ detection ($F_c$) for individual sources, predicted from a Monte Carlo simulation. 
Open blue squares and red circles represent covering fractions for $u$-selected and $g$-selected quasar nebulae, respectively, based on typical SB$_\mathrm{Ly\alpha,z=2.3}$ radial profiles shown in the top panel, where Ly$\alpha$ emissions are embedded into the Ly$\alpha$-band images.
}
\label{fig8}
\end{figure}

Fig.~\ref{fig8} illustrates median values and variations (68th percentiles) in radial profiles of Ly$\alpha$ surface brightness for quasar nebulae at two redshift bins.
For reference, we also show the typical surface brightness profiles at a radius of $>20$ pkpc from previous IFU surveys \citep{ArrigoniBattaia2019,Cai2019}. 
We here adopt the best exponential fit (SB$_\mathrm{Ly\alpha,z=2.3}=13.8\times10^{-17}\exp(-r/15.2)$) to 38 radio-quiet quasars at $z\sim3.17$ based on VLT/MUSE in \citet[table~4]{ArrigoniBattaia2019}. 
Conversely, we scale their best-fit relation by $1/3.69$ to fit the median profile for 17 radio-quiet quasars at $z\sim2.25$ taken from Keck/KCWI in \citet{Cai2019}, as they do not show the best-fit function.
As a result, the $u$-selected quasar nebulae at $z\sim2.14$ show consistent Ly$\alpha$ surface brightness with that at $z\sim2.25$ (Fig.~\ref{fig8}).
The $g$-selected quasar nebulae at $z\sim2.78$ have Ly$\alpha$ surface brightness higher than those of the $u$-selected quasar nebulae and also slightly lower (or higher) than those of quasars at $z\sim3.17$ (or $z\sim2.25$) in previous IFU studies.
Such differentials across quasars at different redshifts may be explained by the increasing trend of Ly$\alpha$ surface brightness from $z\sim2$ to $z\sim3$ \citep{Cai2019}, which is thought to be due to increasing covering factors of Ly$\alpha$-emitting clouds.
However, notably, their results are based on UV-luminous quasars in the range of absolute $i$-band magnitude $-29.7\lesssim M_i[z=2]\lesssim-27.0$, which is systematically brighter than those of our sample with $-27.7<M_i[z=2]<-24.9$ (or $M_i[z=2]=-26.5$ in the median value, see Fig.~\ref{fig7}).
For reference, \citet{Mackenzie2021} have examined Ly$\alpha$ radial profiles for less luminous quasars with $M_i[z=2]\gtrsim-26$ compared to those in \citet[$\sim30$ times brighter on average]{Borisova2016}, and have reported that their Ly$\alpha$ nebulae are four times fainter in mean Ly$\alpha$ surface brightness than bright quasar nebulae previously investigated.
Such a modest correlation between between Ly$\alpha$ surface brightness and quasar luminosity suggests that there may be some sampling bias in the comparison result.

Previous studies have reported that the Ly$\alpha$ surface brightness profile can be well fitted by exponential functions, SB$_\mathrm{Ly\alpha,z=2.3}=C\exp(-r/r_h)$ $\times10^{-17}$ erg~s$^{-1}$cm$^{-2}$arcsec$^{-2}$ \citep{Steidel2011,ArrigoniBattaia2019}.
Therefore, we carry out exponential fit to the Ly$\alpha$ surface brightness profiles using {\tt lmfit} \citep{Newville2014}.
We here mask the central regions within $\sim20$ pkpc radii to avoid significant contributions and complex uncertainties near the centre (Fig.~\ref{fig8}).
Consequently, we obtain the best-fit parameters of $C=2.99\pm3.29$ in the $u$-selected quasar nebulae and $C=9.73\pm3.55$ in the $g$-selected quasar nebulae.
The scale lengths $r_h$ are also estimated to be $r_h=16.60\pm7.05$ and $14.94\pm2.03$ pkpc, respectively, which are in excellent agreement with $r_h=$ 15--16 pkpc in quasars at $z\sim3.17$ based on the VLT/MUSE IFU observation \citep{ArrigoniBattaia2019}.

\begin{figure}
\centering
\includegraphics[width=8cm]{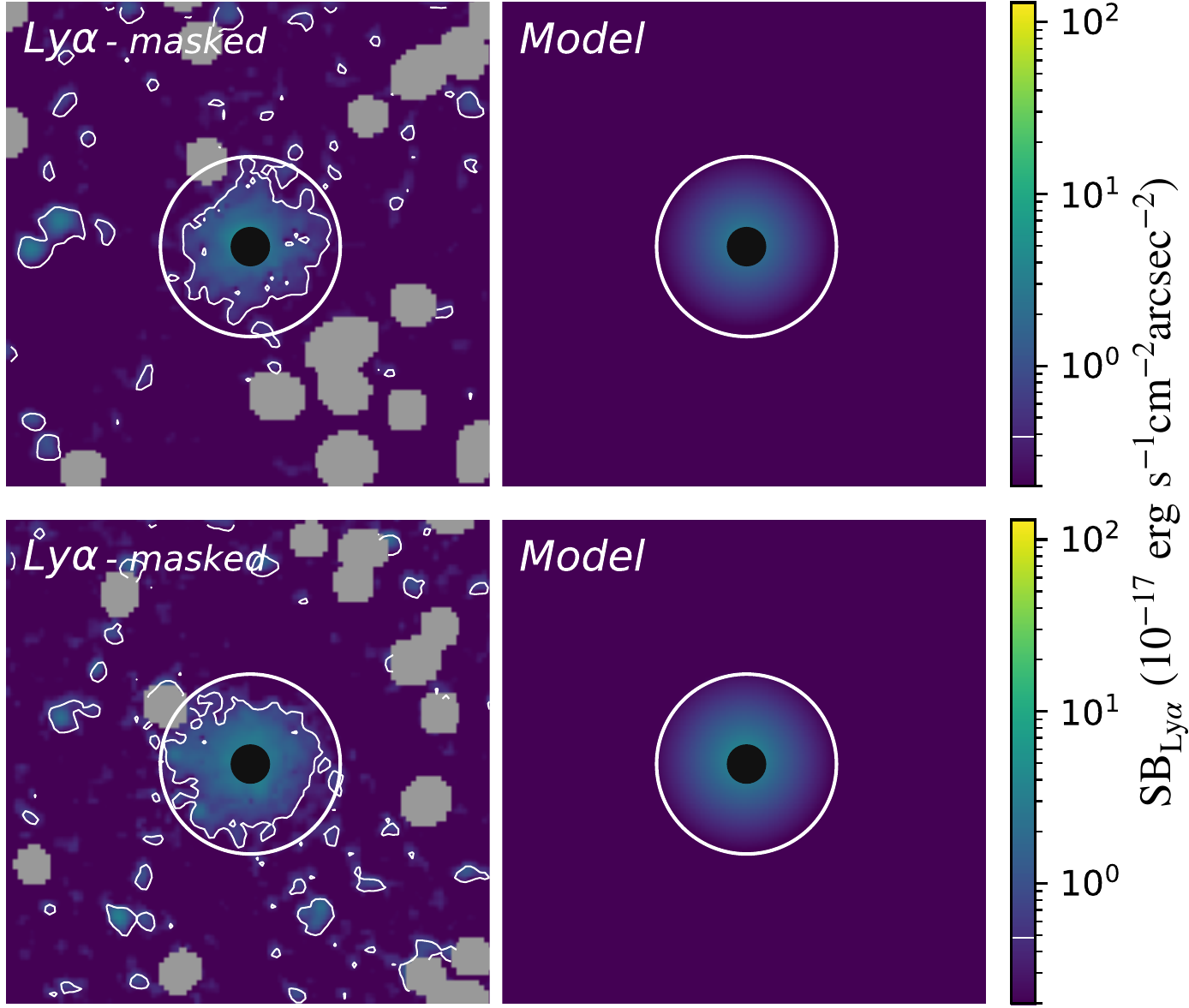}
\caption{
Same as Fig.~\ref{fig5}, but for mock Ly$\alpha$ images using the cutouts of spec-$z$ sources at the same redshifts in the Monte Carlo simulation (Section~\ref{s42}). 
The upper and lower panels show mock examples in the $u$- and $g$-band selections, respectively.
The right panels indicate Ly$\alpha$ emission models (best-fit exponential in each selection) embedded into the Ly$\alpha$-bands, i.e., $u$ or $g$-band.
The typical recovery rates of Ly$\alpha$ detection are shown in Fig.~\ref{fig8}.
}
\label{fig9}
\end{figure}

Furthermore, while the radial profile can be estimated with relatively greater SNR as Ly$\alpha$ surface brightness is averaged over each annulus, the actual broad-band selection is determined by pixel areas above the detection limit on the Ly$\alpha$ images (Section~\ref{s41}). 
Thus, it is important to check the covering fraction ($F_c$) of Ly$\alpha$ detection above the threshold to understand the typical Ly$\alpha$ flux loss expected in our broad-band selection technique.
Specifically, based on the derived exponential functions of Ly$\alpha$ surface brightness profiles, we can calculate the typical covering fraction expected in individual quasar nebulae through the Monte Carlo simulation as follows.

We first cutout images of magnitude-limited spec-$z$ sources ($i<25$) at $z=$ 1.90--2.23 for the $u$-band selection or $z=$ 2.34--3.00 for the $g$-band selection (100 images for each). 
Then, we embed spherically symmetric Ly$\alpha$ emission into the Ly$\alpha$-band ($u$ or $g$) following the formula of Eq.~\ref{eq9} and the best-fit exponential (Fig.~\ref{fig8}) and produce Ly$\alpha$ images ($ugr$ or $gri$) in the same manner as for the main samples (Fig.~\ref{fig9}). 
Based on the obtained mock images, we can measure covering fractions of pixels above the detection threshold ($>2\sigma$ in background deviation, $2\times$ {\tt bg\_std}) in each radius bin. 
The resultant covering fractions are presented in the bottom panel of Fig.~\ref{fig8}, showing that Ly$\alpha$ emission can be recovered more than 90\% up to a radius of $r\sim20$ pkpc.
However, the covering fractions gradually decrease to 50\% at $r\sim30$ pkpc, in which the typical surface brightness reaches the detection threshold, and then drop to only 10\% at $r\sim40$ pkpc.
It means that there would be non-negligible detection loss near the thresholds; in fact, the jagged appearance can be observed near the threshold in the mock Ly$\alpha$ images (Fig.~\ref{fig9}).
Meanwhile, through the Monte Carlo simulation, we also confirm that radially averaged profiles of Ly$\alpha$ surface brightness are well recovered within the margin of error at $r<80$ pkpc, suggesting that the derivations of Ly$\alpha$ radial profiles in Fig.~\ref{fig8} work reasonably well at these radii.


\section{Discussion}
\label{s5}

\subsection{Morphology}
\label{s51}

\begin{figure}
\centering
\includegraphics[width=8cm]{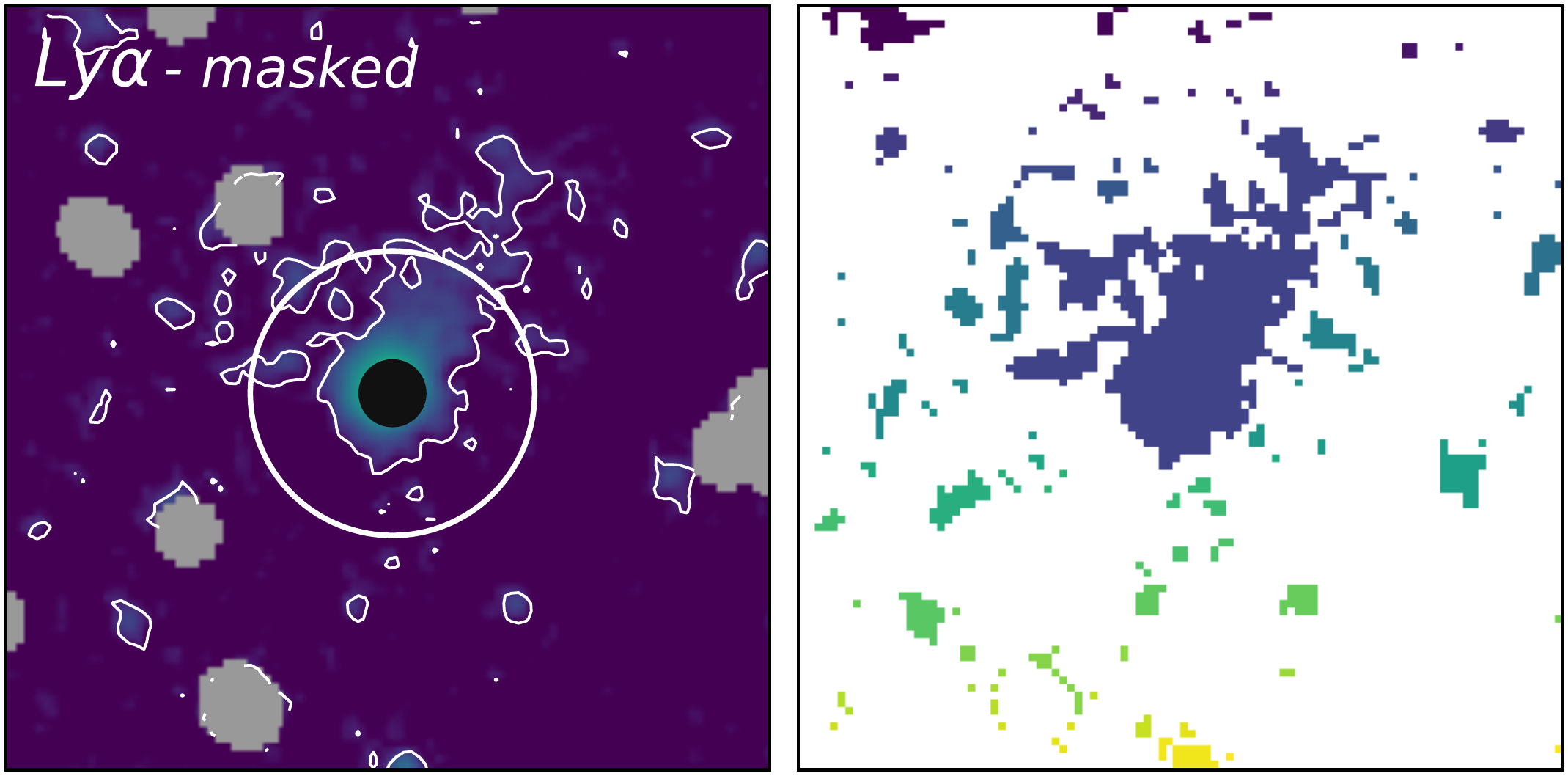}
\caption{
An example of the image segmentation to ({\tt object\_id = 75338966231962053}).
The left panel is a masked Ly$\alpha$ image (same as in Fig.~\ref{fig5}), and the right panel shows the obtained segmentation map.
The white contours indicate $2\sigma$ isophote ($2\times$ {\tt bg\_std}).
Different segmented components are highlighted by different colours. 
}
\label{fig10}
\end{figure}

We discuss Ly$\alpha$ morphology of 24 quasar nebulae in this section. 
To this end, we perform the image segmentation for the Ly$\alpha$ images to separate Ly$\alpha$ nebulae hosted by quasars from residual components produced by different colour terms of projected neighbours (Fig.~\ref{fig3}).
This work defines a connected area within $2\sigma$ isophote ($2\times$ {\tt bg\_std}) around the quasar as a Ly$\alpha$ nebula for use in the morphological analysis.
We adopt the {\tt OpenCV} package (version 4.3.0; \citealt{Bradski2000}) to apply the image segmentation to the individual Ly$\alpha$ images.
Fig.~\ref{fig10} illustrates an example of the resultant segmentation map for one of the most extended Ly$\alpha$ nebulae in our sample ({\tt object\_id}=75338966231962053).

Subsequently, we measure flux-weighted second-order moments ($M_{xx}$, $M_{yy}$, $M_{xy}$) following \citet{ArrigoniBattaia2019},
\begin{eqnarray}
M_{xx} &=& \ev{\frac{(x-x_\mathrm{cen})^2}{r^2}}_f, \nonumber \\
M_{yy} &=& \ev{\frac{(y-y_\mathrm{cen})^2}{r^2}}_f, \\
M_{xy} &=& \ev{\frac{(x-x_\mathrm{cen})(y-y_\mathrm{cen})}{r^2}}_f, \nonumber
\end{eqnarray}
where $(x,y)$, $(x_\mathrm{cen},y_\mathrm{cen})$, and $r$ are pixel coordinates, central coordinates, and radii at $(x,y)$ from the centre in the Ly$\alpha$ images, respectively.
Each moment calculation is weighted by Ly$\alpha$ surface brightness to reduce the data variance, where masked areas in the centres and projected neighbours are excluded (Section~\ref{s3}).
Flux weighting also helps to reduce the effect of low covering fraction of Ly$\alpha$ detection near the edge of Ly$\alpha$ emission, as demonstrated in Section~\ref{s42}. 
Asymmetry parameters ($\alpha$) defined by the minor-to-major axis ratio ($b/a$) can be derived from the second moments,
\begin{eqnarray}
Q\equiv M_{xx}-M_{yy}, \ \ U\equiv 2M_{xy}\\
\alpha = b/a = \frac{1-\sqrt{Q^2+U^2}}{1+\sqrt{Q^2+U^2}},
\end{eqnarray}
where $Q$ and $U$ are the Stokes parameters.
We also measure the projected maximum extent of quasar nebulae based on a segmented isophotal area of the Ly$\alpha$ images (Fig.~\ref{fig10}).

\begin{figure*}
\centering
\includegraphics[width=17cm]{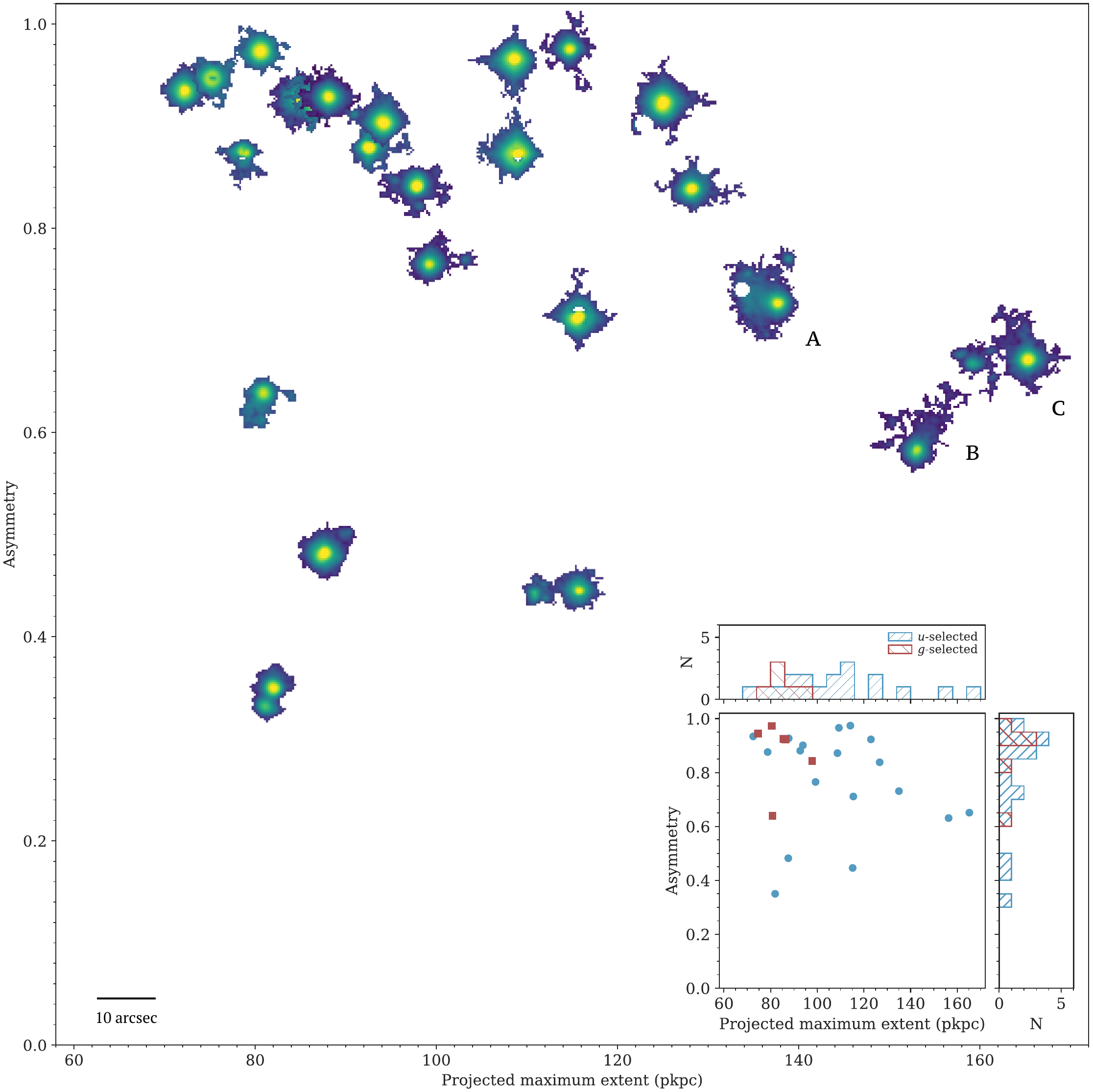}
\caption{
Flux-weighted asymmetry versus projected maximum extent for 24 quasar nebulae at $z=$ 1.9--3.0 with effective Ly$\alpha$ areas $>40$ arcsec$^2$.
The inset panel in the lower right corner shows the comparison of Ly$\alpha$ morphology between the $u$-selected samples at $z=$ 1.90--2.23 (blue symbols and histograms) and $g$-selected samples at $z=$ 2.34--3.00 (red symbols and histograms).
}
\label{fig11}
\end{figure*}

Derived asymmetry and projected maximum extent of the 24 quasar nebulae are summarised in Fig.~\ref{fig11}.
We here highlight three promising quasar nebulae as denoted by A, B, and C in the figure, which show remarkable morphological features with large Ly$\alpha$ effective areas ({\tt Area\_eff} $\gtrsim50$ arcsec$^2$), projected maximum extent ($>130$ pkpc), and asymmetric appearances.
These unique structures evoke the ELANe discovered to date \citep{Cantalupo2014,Hennawi2015,Cai2017,ArrigoniBattaia2018,Li2024}.
We should note that our samples have fainter and smaller Ly$\alpha$ nebulosities than these ELANe with the projected maximum extent beyond 280 pkpc, whereas they sufficiently meet the definition of Type I ELAN ($>100$ pkpc with $M_\mathrm{UV}<-22$) by \citet{Li2024}.
Other than these, there are ten quasar nebulae with a maximum extent of $\sim100$ pkpc; however, some of them may be overestimated due to diffraction spikes caused by the central bright quasars. 
Three quasar nebulae with $\alpha<0.5$ tend to have bright companions in the Ly$\alpha$ images, which suggests that their high asymmetry is primarily because of bright Ly$\alpha$ emitter candidates in the vicinity, as illustrated in Fig.~\ref{fig11}. 

We also check if there is a redshift dependence in asymmetry of Ly$\alpha$ nebulosities for the quasar nebulae (Fig.~\ref{fig11}).
\citet{Cai2019} have reported that, combined with the sample from \citet{ArrigoniBattaia2019}, possible morphological evolution of quasar nebulae from $z\sim3$ to $z\sim2$ in the sense that Ly$\alpha$ morphology may tend to be more asymmetric and clumpy at lower redshift owing to the decreasing covering factor and overall mass of Ly$\alpha$-emitting clouds.
However, no significant difference in asymmetry parameter ($\alpha$) has been detected between the $u$-selected and $g$-selected quasar nebulae based on our tentative measurements depicted in the bottom right of Fig.~\ref{fig11}.
We note that our Ly$\alpha$ imaging is significantly shallower than the IFU surveys by \citet{ArrigoniBattaia2019,Cai2019}, which means that our measurements do not take into account the outer components as much as the previous studies. 
It should be also noted that the redshift dimming-corrected surface brightness depths are different between the $u$ and $g$-band selections (a factor of 3.5) and the sample sizes are limited to only 18 and 6 quasar nebulae, respectively.
Furthermore, as mentioned in Section~\ref{s42}, our samples are less luminous quasars ($-27.7<M_i[z=2]<-24.9$) than those in the previous studies ($-29.7\lesssim M_i[z=2]\lesssim-27.0$).
Given all these limitations, we conclude that the lack of redshift evolution of asymmetric natures in our samples would be simply sampling effects and is difficult to view as an intrinsic trend at this moment.

\subsection{Environmental dependence}
\label{s52}

Despite considerable surveys of high redshift quasars, environmental dependence of quasars on galaxy overdensities and properties remain controversial (e.g., \citealt{Venemans2007,Miley2008,Galametz2012,Hatch2014,Jones2015,Uchiyama2018,Uchiyama2020,Suzuki2024}).
This study examines the environmental dependence of quasar nebulae by measuring local overdensities around quasars and those with large Ly$\alpha$ nebulae using photometric redshifts \citep{Tanaka2018}.
To investigate the environmental dependence, we construct three samples in the HCD-JF: (1) all quasars but exclusive of shallow images with the pixel background deviation above $0.67\times10^{-17}$ erg~s$^{-1}$cm$^{-2}$arcsec$^{-2}$ (see Section~\ref{s41}), (2) quasars with large Ly$\alpha$ nebulae ranking in the top 16th-percentiles given the redshift, which are located above the threshold in the lower panel of Fig.~\ref{fig6}, and (3) the magnitude-limited control sample at $z=$ 1.9--3.0. 
The sample sizes are $N=$ 409, 69, and 3228 across the survey field, respectively.
We apply the $i$-band magnitude cut ($i<22$ mag that comprises 95\% of our quasar sample) for the control sample at $z=$ 1.9--3.0 for a relatively fair comparison rather than using random positions or all spec-$z$ sources, including faint galaxies; this is because more luminous galaxies regardless of black hole activities tend to host more massive haloes and contain a greater number of satellites (e.g., \citealt{Zehavi2005,Zehavi2011,Coil2008}); hence, they would be located in higher densities when compared to random sources or fields.
Otherwise, we cannot ensure whether an obtained trend is driven by quasar environments or the clustering effects of brighter objects.

\begin{figure*}
\centering
\includegraphics[width=12cm]{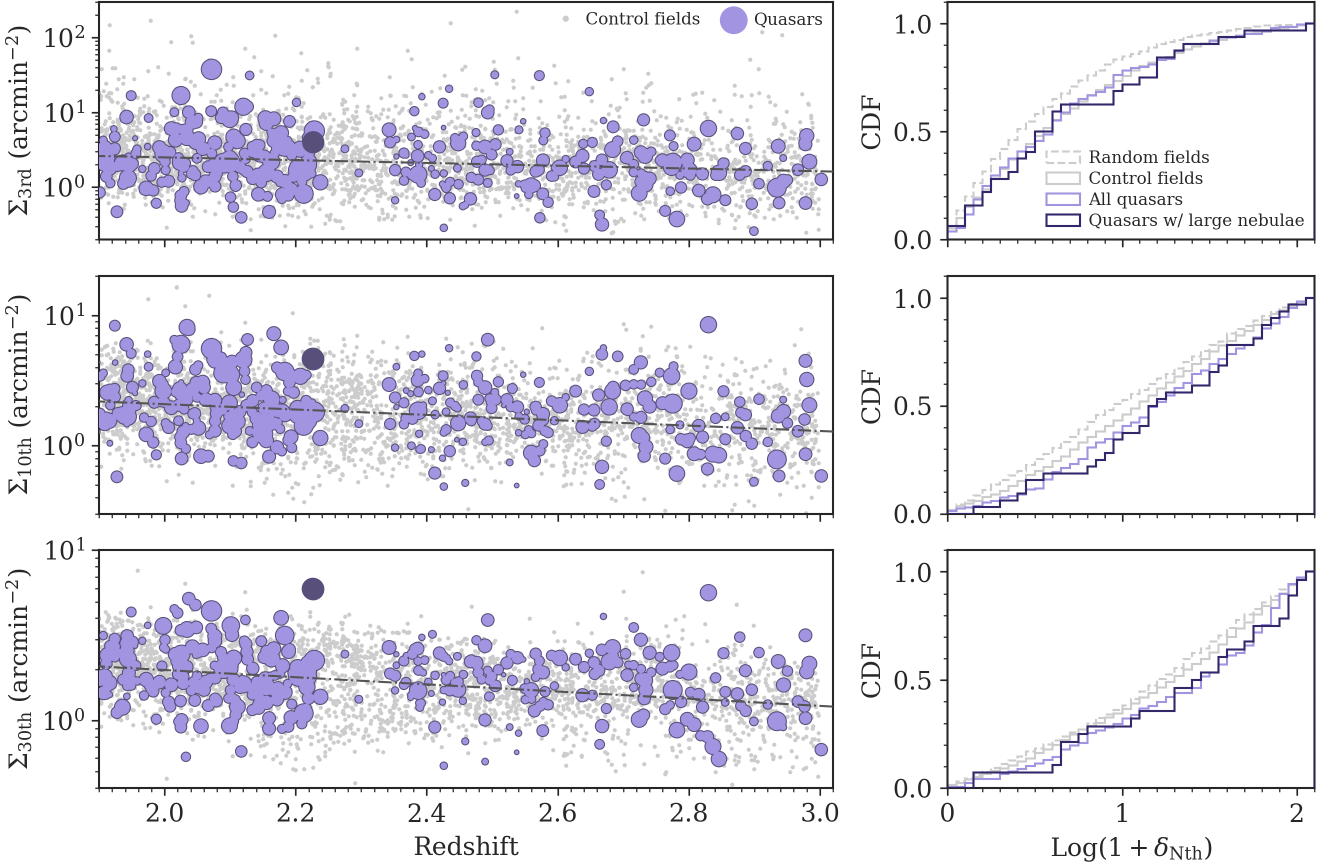}
\caption{
(Left) $N$-th nearest neighbour densities for quasars and control samples, shown for $N=3,10,30$ (first to third rows, respectively). 
Purple symbols represent 409 quasars, with symbol sizes scaled by effective Ly$\alpha$ area ({\tt Area\_eff}).
Grey dots show $i$-band magnitude limited ($i<22$ mag) control samples with spectroscopic redshifts.
The most extended nebula among our targets is highlighted in dark purple ({\tt object\_id} = 37484563299081621 at $z=2.226$).
(Right) CDFs of overdensities, $\log(1+\delta_\mathrm{Nth})$ for all quasars (light purple), quasars with large nebulae in the top 16th-percentile (purple), and control samples (grey). 
No systematic differences are detected in any comparison, according to the KS test ($p>0.1$).
Overdensity distributions for 4000 random blank fields are also shown as grey dashed histograms, skewing toward lower values than the three samples ($p\ll0.01$).
}
\label{fig12}
\end{figure*}

For each object in the three samples, we estimate the $N$-th nearest neighbour densities ($\Sigma_\mathrm{Nth} = N/\pi d_\mathrm{Nth}^2$ arcmin$^{-2}$) based on photometric redshifts of $i$-band magnitude limited ($i<25$) sources (see Section~\ref{s3}).
We here adopt three $N$ values ($N=3,10,30$) to check whether or not the outcomes depend on the choice of $N$. 
We count the nearest neighbours within the redshift range of $\delta z=(z_\mathrm{neighbour}-z_\mathrm{spec,target})/(1+z_\mathrm{spec,target})\pm0.05$ from each target, including the target itself, and then determine the $N$-th neighbour distance ($d_\mathrm{Nth}$).
Notably, small $d_\mathrm{Nth}$ values would be subject to the projection effect of foreground and background sources in light of the high outlier rate of 0.3--0.4 at $|\delta z|>0.05$ in the photo-$z$ estimation (Eq.~\ref{eq3}).
The obtained $\Sigma_\mathrm{Nth}$ overdensities are summarised in the left panel of Fig.~\ref{fig12}.
Furthermore, we perform a linear fit to the $\Sigma_\mathrm{Nth}$--$z$ distributions based on the control sample with {\tt lmfit} \citep{Newville2014} to derive the mean local densities at a given redshift, $\overline{\Sigma}_\mathrm{Nth}$ as shown in Fig.~\ref{fig12}.
Based on the mean overdensities, we calculate delta variation for each $\Sigma_\mathrm{Nth}$ including correction of redshift dependence of the density measurement ($\delta_\mathrm{Nth}=(\Sigma_\mathrm{Nth}-\overline{\Sigma}_\mathrm{Nth})/\overline{\Sigma}_\mathrm{Nth}$).

The cumulative distribution functions (CDFs) of the local overdensities $\delta_\mathrm{Nth}$ are represented on the right in Fig.~\ref{fig12}.
Overall, we do not detect any significant difference between the quasar fields and the control fields given $p$-value $>0.1$ in the two-sample Kolmogorov–Smirnov (KS) test in each comparison even though they show systematically higher densities compared to the 4000 blank fields randomly selected from the entire survey area ($p\ll0.01$). 
We thus conclude that there is no environmental dependence on quasar activities in our current data.
The result seems reasonable given that the quasar lifetimes are sufficiently short (e.g., $\lesssim$ a few Myr, according to \citealt{Khrykin2021,Morey2021}) compared to the hierarchical structure formation timescales. 
We here note that all selected fields are systematically denser when we compare those with 4000 random blank fields (Fig.~\ref{fig12}).
Also, there have been predictions and reports that the number density of low-mass star-forming galaxies decreases in the vicinity of bright quasars due to photo-evaporation \citep{Benson2002,Kashikawa2007,Kikuta2017,Uchiyama2019,Suzuki2024}. 
However, it is outside of the scope of this paper as our magnitude-limited selection does not cover such faint galaxies.
Moreover, it is important to note that our density measurement may not trace the large-scale structures enough to examine the local environmental dependence due to the photometric redshift errors ($\delta z\sim0.05$).
The forthcoming strategic program with the Subaru Prime Focus Spectrograph will provide a large spec-$z$ sample and underlying Ly$\alpha$ opacity with IGM tomography in the HSC-JF, enabling us to better address these questions \citep{Takada2014}.

Owing to previous efforts, it seems that we have almost reached a consensus that large Ly$\alpha$ nebulae tend to be located in high-density regions such as galaxy protoclusters and filaments \citep{Matsuda2004,Nowotka2022,ArrigoniBattaia2023,Ramakrishnan2023,Li2024,Pensabene2024,Zhang2025b}.
However, when we compare the number density distributions between quasar fields with and without the relatively large Ly$\alpha$ nebulae and the control fields, we do not detect a statistically significant difference between the three samples (Fig.~\ref{fig12}). 
This seems to conflict with previous results; whereas, it is difficult to judge it because they calculate overdensities in different manners and sample selections.
To give an important point, the most massive halos that would seed galaxy clusters at $z=$ 2--3 have already reached more than $1\times10^{13}$ M$_\odot$, which is well above the quasar sweet spot around $\sim2\times10^{12}$ M$_\odot$ suggested by \citet[see also, e.g., \citealt{White2012,Rodriguez-Torres2017,Massingill2024}]{Zhang2025}.
Thus, it may not be surprising that we do not see such a marked environmental dependence at this redshift range.

\begin{figure*}
\centering
\includegraphics[width=12cm]{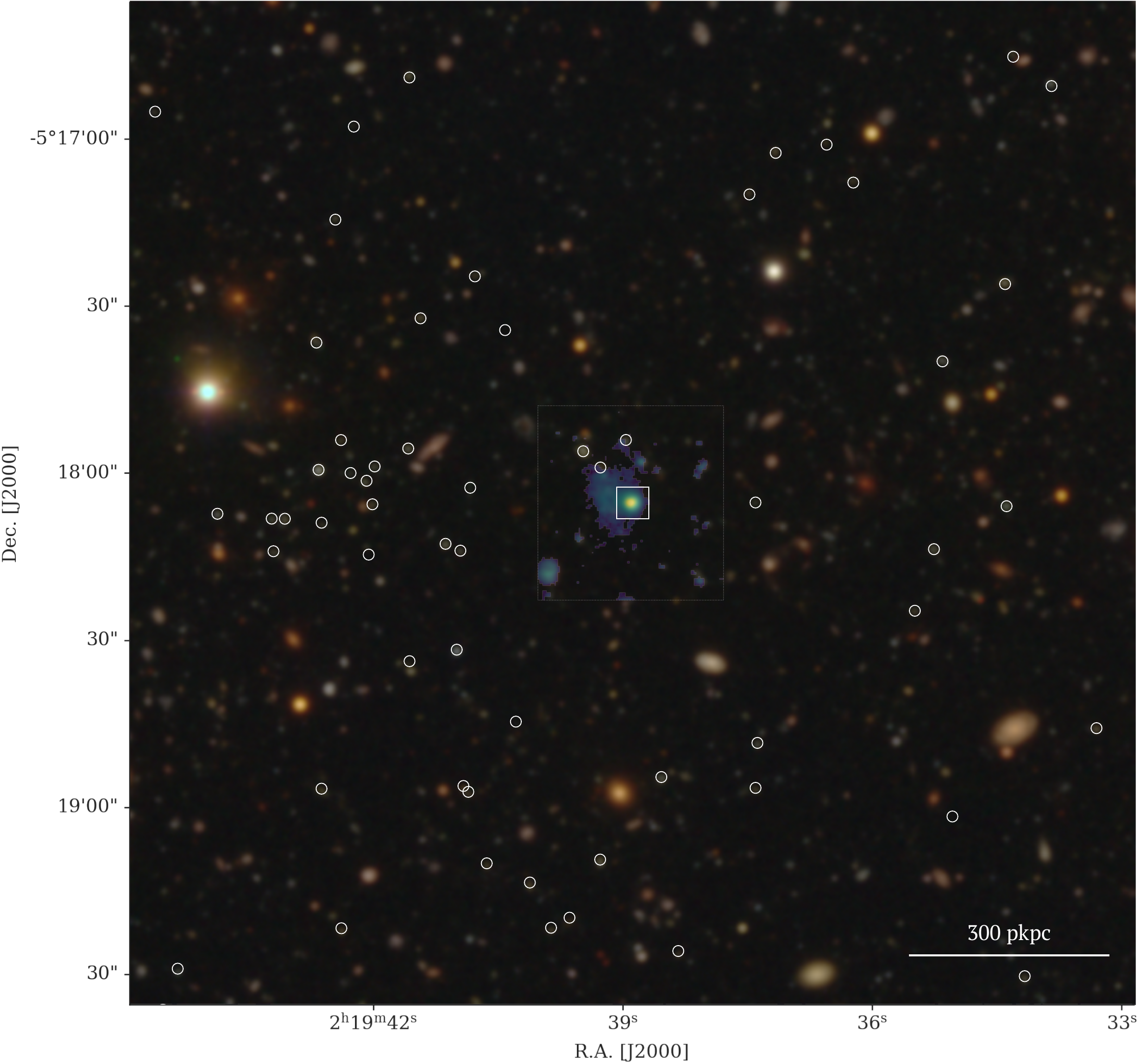}
\caption{
On-sky distribution of photo-$z$ selected neighbours within $\delta z\pm0.05$ (white open circles) surrounding the most extended quasar nebula in our sample (the white open square, {\tt object\_id} = 37484563299081621 at $z=2.226$).
The background image is an RGB cutout based on the $u^\ast gr$ bands ($3\times3$ arcmin$^2$), overlaid with the Ly$\alpha$ nebula image around the central quasar (Fig.~\ref{fig5}).
}
\label{fig13}
\end{figure*}

It is important to note that \citet{Nowotka2022} have reported the ubiquitous overdensities around four known ELANe at $z>2$ based on SCUBA-2 850 $\mu$m imaging (see also \citealt{ArrigoniBattaia2023}).
We consider that such enormous Ly$\alpha$ nebulae extending beyond a CGM scale would be unique exceptions as they require vast cool gas that can be associated only with massive protocluster haloes.
In our sample, the largest Ly$\alpha$ nebula at $z=2.226$ ({\tt object\_id} = 37484563299081621 or the nebula-`A' in Fig.~\ref{fig11}) seems to be located in a high-density region, while other large nebulae are not all in high-density regions, and rather they are distributed over a wide range of densities (Fig.~\ref{fig12}). 
Fig.~\ref{fig13} shows the spatial distribution of galaxy candidates associated with the nebula-A at $z=2.226$. 
An apparent density peak is observed in a region approximately 400 pkpc east of the nebula; however, due to the absence of spec-$z$ sources (Section~\ref{s3}), further observations are needed to confirm this protocluster candidate.
The spatial offset from the density peak may also indicate that the nebula-A does not belong to the most massive halo of this high-density region.
Additionally, we observe a quasar nebula ({\tt object\_id} = 41624100023598926) showing a clear overdensity at $z=2.82$ in the $g$-band selection.
Although the effective Ly$\alpha$ area of this nebula is in the top $\sim3\%$ of the whole quasar sample at $z=2.34$--3.00, it is almost average in effective Ly$\alpha$ area among the $g$-band selected nebulae and no other features of note are observed.

\section{Conclusions}
\label{s6}

We search for extended Ly$\alpha$ nebulae around 483 SDSS/eBOSS quasars at $z=$ 1.9--3.0 in the HCD-JF over 13 deg$^2$, as a RIDEN pilot survey.
Although such a systematic search beyond a ten deg$^2$ has been very challenging until the series of our studies, the broad-band selection using the deep wide-field $ugri$ data provided by HSC-SSP and CLAUDS ($>26$ mag; \citealt{Aihara2019,Sawicki2019}) has made it possible down to the detection limit below SB$_\mathrm{Ly\alpha}=1\times10^{-17}$ erg~s$^{-1}$cm$^{-2}$arcsec$^{-2}$ in 1 arcsec$^2$.
This work additionally adopts $u$-band selection from the previous study \citep{Shimakawa2022}, which can go deeper than the $g$-band selection because of its $\sim2$ times narrower filter width and lower redshifts, i.e., lower redshift dimming of the corresponding targets.
Here, the accumulation of enormous observing time plus archive data in CFHT reaching 462 hours \citep{Sawicki2019} greatly helps us to dive into Ly$\alpha$ surface brightness down to $\sim4\times10^{-18}$ erg~s$^{-1}$cm$^{-2}$arcsec$^{-2}$ in 1 arcsec$^2$ aperture.
Photometric redshifts based on multi-band photometry from $u$ to $K$ -band allow us to mask projected neighbours more effectively and investigate local environments around hundreds of quasars with and without large Ly$\alpha$ nebulae for the first time.
The major highlights in this work are organised as follows:
\begin{description}

\item[---] 
Based on the $ugr$ and $gri$ broad-band selections, we obtain 39 and 17 quasars at $z=$ 1.90--2.23 (or median $z=2.10$) and $z=$ 2.34--3.00 (or median $z=2.77$), respectively, hosting relatively large Ly$\alpha$ nebulae with effective areas $>40$ arcsec$^2$.
When considering cosmological redshift dimming, quasar nebulae at $z\sim2.14$ tend to have lower Ly$\alpha$ surface brightness than those at $z\sim2.78$, showing a good agreement with a previous study \citep{ArrigoniBattaia2019,Cai2019}.
Obtained spatial properties of Ly$\alpha$ emission, such as the effective areas and projected maximum extent, are available as online material (Table~\ref{tab3}).

\item[---] 
We investigate and compare Ly$\alpha$ morphology for 24 selected quasar nebulae by dividing the sample into two groups by the colour selection ($ugr$ and $gri$).
We find three promising candidates of giant Ly$\alpha$ nebulae that exhibit large effective areas, projected maximum extents beyond 130 pkpc, and asymmetric morphology. 
However, we do not detect a clear difference in asymmetry distributions between quasar nebulae at $z\sim2.10$ and $z\sim2.77$. 
Note that our survey depth is significantly shallower than previous IFU studies, which may lead to the apparent similarity in symmetries of Ly$\alpha$ nebulae between the two redshift slices.

\item[---]
We study the environmental dependence of quasar nebulae by comparing them with control samples using photometric redshifts.
Consequently, we do not see a clear dependence of quasars on galaxy overdensities, regardless of the sizes of their Ly$\alpha$ nebulosities.
Non-detection of environmental dependence seems to conflict with previous studies; however, we should note that there are differences in the way that each comparison is made.
In particular, our sample would not contain ELANe extending hundreds pkpc in scale; therefore, the statistical trend observed in this study may not be applicable for such an extreme case \citep{Nowotka2022}. 
Indeed, the largest Ly$\alpha$ nebula in our sample seems to be positioned in a high-density region.
We are also concerned that the current density measurement may not reach a sufficient resolution that allows us to constrain the local environmental dependence given the photo-$z$ uncertainties.

\end{description}

A follow-up IFU survey is crucial for validating the surface brightness measurement and improving the selection methodology.
Furthermore, the upcoming wide-field IGM tomography with the Subaru Prime Focus Spectrograph will mostly cover the survey field of this work \citep{Takada2014}.
Thus, it will help us delve into the environmental dependence of large Ly$\alpha$ nebulae in a less biased way since the neutral hydrogen can trace underlying large-scale structures independent of galaxy selection bias \citep{Lee2014,Lee2016,Lee2018}, although this technique may underestimate the highest densities affected by AGN preheating \citep{Pratt2010,Chaudhuri2013,Iqbal2017,Kooistra2022}.


\section{Future prospects}
\label{s7}

\begin{figure*}
\centering
\includegraphics[width=14cm]{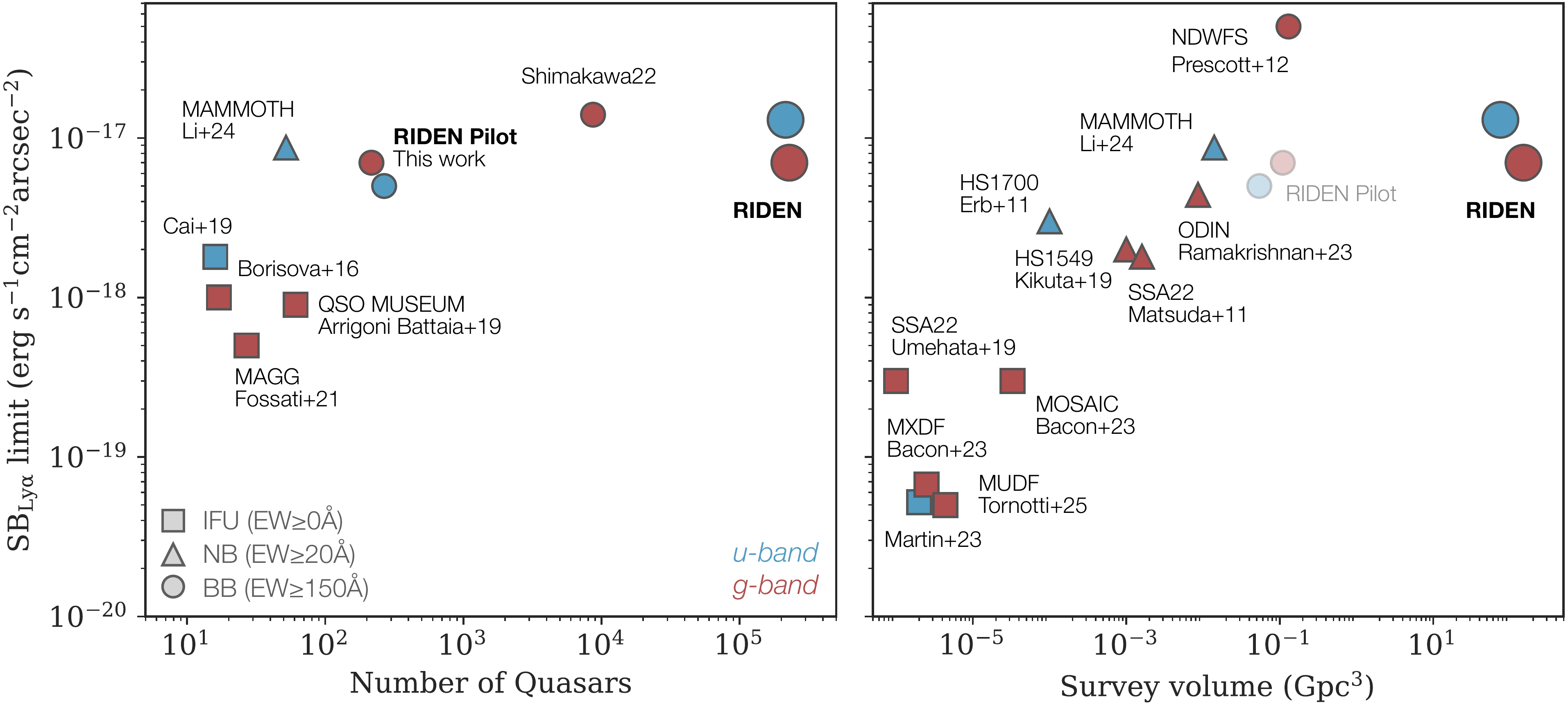}
\caption{
Ly$\alpha$ surface brightness ($\mathrm{SB_{Ly\alpha}}$) limit versus (left) the number of quasars and (right) the survey volume in various survey programs as given in the figures \citep{Erb2011,Matsuda2011,Prescott2012,Borisova2016,ArrigoniBattaia2019,Cai2019,Kikuta2019,Umehata2019,Fossati2021,Shimakawa2022,Bacon2023,Martin2023,Ramakrishnan2023,Li2024,Tornotti2025}.
While those in this work are termed as ``RIDEN Pilot'' here, hypothetical cases if we apply the broad-band selection to the entire field are denoted in the right panel.
The $\mathrm{SB_{Ly\alpha}}$ limit is defined as 2$\sigma$ detection limit in 1 arcsec$^2$ aperture, which has been commonly used as a survey depth.
The square, triangle, and circle symbols depict different observation types, i.e., IFU spectroscopy, narrow-band imaging, and broad-band imaging, respectively, where the former technique has better detectability in terms of EW$_\mathrm{Ly\alpha}$.
The selections in the $u$-band or corresponding wavelength are shown by the blue colours and those in $g$-band are represented by the reds.
We note that some of the data points, particularly those in the right panel, are our rough estimates based on values found in the literature (i.e., not based on a homogeneous definition) because they cannot be uniquely determined (see Appendix~\ref{a1}).
}
\label{fig14}
\end{figure*}

Our results highly encourage an extension of the project to the forthcoming Rubin/LSST \citep{Ivezic2019}. 
Fig.~\ref{fig14} summarises $2\sigma$ limits of Ly$\alpha$ surface brightness, mostly in 1 arcsec$^2$ aperture, in the previous studies of Ly$\alpha$ nebulae and/or diffuse Ly$\alpha$ filaments. 
One should note here that our rough estimates are included in this figure due to inhomogeneous definitions of the limiting depth depending on the scientific objectives and limited information in the literature (see Appendix~\ref{a1}).
This comparative diagram well demonstrates the trade-off between the survey depth and the number of quasars or survey volume: IFU spectroscopy typically probes smaller fields compared to imaging surveys but achieve greater depths \citep{Borisova2016,ArrigoniBattaia2019,Cai2019,Umehata2019,Fossati2021,Bacon2023}, particularly with modern instruments on the large aperture telescopes such as VLT/MUSE \citep{Bacon2010} and Keck/KCWI \citep{Morrissey2018}.
Narrow-band imaging has long been used and a powerful method to detect extended Ly$\alpha$ nebulae and Ly$\alpha$ emitters \citep{Erb2011,Matsuda2011,Kikuta2019,Ramakrishnan2023,Li2024}. 
It has a wider field-of-view especially when using Subaru/HSC \citep{Miyazaki2018} and also allows to find the large-scale structure using Ly$\alpha$ emitters as indicators, albeit with shallower depth compared to IFU survey due to the limitation of filter bandwidth. 

The broad-band selection is a more extreme approach of the traditional narrow-band selection, which provides an overwhelming survey volume at the expense of depth and detectability of low EW$_\mathrm{Ly\alpha}$ \citep{Prescott2012,Shimakawa2022}.
Therefore, the disadvantage of this technique is that it is realistically limited to detecting bright Ly$\alpha$ nebulae with high EW$_\mathrm{Ly\alpha}$ ($\gtrsim150$ \AA) that are generally hosted by luminous quasars as presented by this paper. 
Notably, Rubin/LSST will perform deep $ugrizy$-band imaging over 18,000 deg$^2$, enabling us to search for bright Ly$\alpha$ nebulae across the unprecedentedly wide field of $\sim15$ and $\sim1400$ times wider than the HSC-SSP Wide Layer \citep{Shimakawa2022} and this pilot survey.

However, several technical issues can be identified for practical use. 
For example, given the enormous data volume ($>100$ PB) of Rubin/LSST, it is not time-efficient to apply the broad-band selection directly to the entire field, and rather, it would be practical to pre-select quasar or protocluster fields. 
Toward that end, building a quasar sample is quite important since the current largest dataset by the SDSS/BOSS quasar survey is mostly outside the Rubin/LSST footprint. 
Here, the planned multi-wavelength and cadence analyses may help to select promising target fields \citep{Ivezic2019,Bianco2022,Li2022}.
At this moment, the numbers of quasars covered by RIDEN in Fig.~\ref{fig14} are provisional, which are simply extrapolated from the quasar number densities in the DR16Q catalogue (\citealt{Lyke2020}, see Appendix~\ref{a1}). 
These challenges will need to be overcome in the future, e.g., by making better use of Rubin/LSST mock data and early releases, and by coordinating with other wide-field imaging and spectroscopic surveys.

\section*{Acknowledgements}

We thank the anonymous referee for useful comments. 
This research is based on data collected at Subaru Telescope, which is operated by the National Astronomical Observatory of Japan.
We are honoured and grateful for the opportunity of observing the Universe from Maunakea, which has the cultural, historical and natural significance in Hawaii.
RS acknowledges financial supports from Waseda University Grant for Special Research Projects (2023C-590 and 2024R-057) and MEXT/JSPS KAKENHI Grant Numbers (22K14078 and 25K01044).
SK acknowledges financial supports from MEXT/JSPS KAKENHI Grant Numbers (24KJ0058 and 24K17101).
HK acknowledges financial supports from MEXT/JSPS KAKENHI Grant Numbers (23KJ2148 and 25K17444).

The Hyper Suprime-Cam (HSC) collaboration includes the astronomical communities of Japan and Taiwan, and Princeton University. The HSC instrumentation and software were developed by the National Astronomical Observatory of Japan (NAOJ), the Kavli Institute for the Physics and Mathematics of the Universe (Kavli IPMU), the University of Tokyo, the High Energy Accelerator Research Organization (KEK), the Academia Sinica Institute for Astronomy and Astrophysics in Taiwan (ASIAA), and Princeton University. Funding was contributed by the FIRST program from Japanese Cabinet Office, the Ministry of Education, Culture, Sports, Science and Technology (MEXT), the Japan Society for the Promotion of Science (JSPS), Japan Science and Technology Agency (JST), the Toray Science Foundation, NAOJ, Kavli IPMU, KEK, ASIAA, and Princeton University. 
This paper makes use of software developed for the Large Synoptic Survey Telescope. We thank the LSST Project for making their code available as free software at \url{http://dm.lsst.org}.

The Pan-STARRS1 Surveys (PS1) have been made possible through contributions of the Institute for Astronomy, the University of Hawaii, the Pan-STARRS Project Office, the Max-Planck Society and its participating institutes, the Max Planck Institute for Astronomy, Heidelberg and the Max Planck Institute for Extraterrestrial Physics, Garching, The Johns Hopkins University, Durham University, the University of Edinburgh, Queen’s University Belfast, the Harvard-Smithsonian Center for Astrophysics, the Las Cumbres Observatory Global Telescope Network Incorporated, the National Central University of Taiwan, the Space Telescope Science Institute, the National Aeronautics and Space Administration under Grant No. NNX08AR22G issued through the Planetary Science Division of the NASA Science Mission Directorate, the National Science Foundation under Grant No. AST-1238877, the University of Maryland, and Eotvos Lorand University (ELTE) and the Los Alamos National Laboratory.

These data were obtained and processed as part of the CFHT Large Area U-band Deep Survey (CLAUDS), which is a collaboration between astronomers from Canada, France, and China described in \citet{Sawicki2019}. CLAUDS is based on observations obtained with MegaPrime/ MegaCam, a joint project of CFHT and CEA/DAPNIA, at the CFHT which is operated by the National Research Council (NRC) of Canada, the Institut National des Science de l’Univers of the Centre National de la Recherche Scientifique (CNRS) of France, and the University of Hawaii. CLAUDS uses data obtained in part through the Telescope Access Program (TAP), which has been funded by the National Astronomical Observatories, Chinese Academy of Sciences, and the Special Fund for Astronomy from the Ministry of Finance of China. CLAUDS uses data products from TERAPIX and the Canadian Astronomy Data Centre (CADC) and was carried out using resources from Compute Canada and Canadian Advanced Network For Astrophysical Research (CANFAR).

Funding for the Sloan Digital Sky Survey IV has been provided by the Alfred P. Sloan Foundation, the U.S. Department of Energy Office of Science, and the Participating Institutions. SDSS-IV acknowledges support and resources from the Center for High Performance Computing  at the University of Utah. The SDSS website is \url{www.sdss.org}.

SDSS-IV is managed by the Astrophysical Research Consortium for the Participating Institutions of the SDSS Collaboration including the Brazilian Participation Group, the Carnegie Institution for Science, Carnegie Mellon University, Center for Astrophysics | Harvard \& Smithsonian, the Chilean Participation Group, the French Participation Group, Instituto de Astrof\'isica de Canarias, The Johns Hopkins University, Kavli Institute for the Physics and Mathematics of the Universe (IPMU) / University of Tokyo, the Korean Participation Group, Lawrence Berkeley National Laboratory, Leibniz Institut f\"ur Astrophysik Potsdam (AIP),  Max-Planck-Institut f\"ur Astronomie (MPIA Heidelberg), Max-Planck-Institut f\"ur Astrophysik (MPA Garching), Max-Planck-Institut f\"ur Extraterrestrische Physik (MPE), National Astronomical Observatories of China, New Mexico State University, New York University, University of Notre Dame, Observat\'ario Nacional / MCTI, The Ohio State University, Pennsylvania State University, Shanghai Astronomical Observatory, United Kingdom Participation Group, Universidad Nacional Aut\'onoma de M\'exico, University of Arizona, University of Colorado Boulder, University of Oxford, University of Portsmouth, University of Utah, University of Virginia, University of Washington, University of Wisconsin, Vanderbilt University, and Yale University.

We would like to thank Editage (\url{www.editage.com}) for English language editing.
This work made extensive use of the following tools, {\tt NumPy} \citep{Harris2020}, {\tt Matplotlib} \citep{Hunter2007}, the Tool for OPerations on Catalogues And Tables, {\tt TOPCAT} \citep{Taylor2005}, a community-developed core Python package for Astronomy, {\tt Astopy} \citep{AstropyCollaboration2013}, and Python Data Analysis Library {\tt pandas} \citep{Reback2021}.

\section*{Data Availability}

The data underlying this article are available in part on the public data release site of Hyper Suprime-Cam Subaru Strategic Program (\url{https://hsc.mtk.nao.ac.jp/ssp/data-release/}).
For all other survey data, readers would need to contact the respective coordinators.
The source catalogue and processed images are also accessible as online material.
When one wants to obtain coadd imaging data for specific sources, the authors recommend using the user-friendly cutout tool (\url{https://hsc-release.mtk.nao.ac.jp/das_cutout/pdr3/}) or the interactive sky viewer {\tt hscMap} (\url{https://hsc-release.mtk.nao.ac.jp/hscMap-pdr3/app}).



\bibliographystyle{mnras}
\bibliography{rs22c} 



\appendix

\section{Previous surveys}
\label{a1}

This appendix section provides a summary of previous Ly$\alpha$ narrow-band and IFU surveys for Ly$\alpha$ nebulae at $z=$ 2--4 shown in Fig.~\ref{fig14}, where individual values are summarised in Table~\ref{tab4}.
It should be noted that some of these values listed here are rough estimates due to different definitions of Ly$\alpha$ surface brightness limits and lack of specific information in the literature as explained below.

First, we refer to nine IFU survey programs using VLT/MUSE \citep{Bacon2010} or Keck/KCWI \citep{Morrissey2018}.
IFU observations have increasingly contributed to the detection and characterisation of diffuse Ly$\alpha$ emission since the advent of these modern instruments. 
Three previous IFU studies targeting quasars have corresponding values needed to plot in Fig.~\ref{fig14} (\citealt[section~3.3]{Borisova2016}; \citealt[section~2]{ArrigoniBattaia2019}; \citealt[table~1]{Cai2019}), and thus, we take typical values from them.
On the other hand, we do not find a specific description of the depth for a quasar IFU survey, MAGG \citep{Fossati2021}, and we roughly estimate it to be $0.5\times10^{-18}$ erg~s$^{-1}$cm$^{-2}$arcsec$^{-2}$ based on their typical 4 hour integration per source (section~2), which is approximately four times as long as \citet[section~2.2]{Borisova2016}. 
Ultra-deep IFU surveys aiming to detect diffuse Ly$\alpha$ filaments reach an order of $10^{-19}$ erg~s$^{-1}$cm$^{-2}$arcsec$^{-2}$ or even below by long integrations with VLT/MUSE (\citealt{Umehata2019}; \citealt[fig.~2, section~7]{Bacon2023}; \citealt{Martin2023}; \citealt[fig.~1, section~2.1]{Tornotti2025}).
We note here that their survey volumes shown in Table~\ref{tab4} are our rough order of magnitude estimates, not values specified in the literature, based on the survey areas and the redshift ranges examined; therefore, they can be taken only as guides. 
Additionally, because they have used mosaics with many frames, their depths described in the literature may not correspond one-to-one with the rough estimates. 
It should be also noted that these deep surveys focus on underlying Ly$\alpha$ emission significantly fainter than the quasar nebulae. 
For that reason, except \citet[fig.~9, section~6.1]{Bacon2023}, their sensitivity limits are calculated at different areas from 1 arcsec$^2$ aperture (e.g., \citealt{Umehata2019} have measured the noise level with 4 arcsec diameter aperture, see their supplementary materials for details). 
Besides, the spectral resolution and the number of binned spectral pixels are not homogenous among these IFU surveys.

As for the narrow-band imaging surveys, we have obtained the depths and survey volumes through the literature \citep{Erb2011,Matsuda2011,Ramakrishnan2023} or private communications \citep{Kikuta2019,Li2024}.
Here, \citet{Ramakrishnan2023} have used 3 arcsec$^2$ aperture (section~3.2), while we cannot refer to which size apertures have been adopted for \citet[section~3]{Erb2011} and \citet[table~1]{Matsuda2011}.
We also directly refer to the survey volumes for these studies as they are specifically mentioned in the literature (\citealt[section~3.1]{Erb2011}; \citealt[section~2]{Matsuda2011}; \citealt[section~6]{Ramakrishnan2023}).

As for the broad-band selections, we can take the corresponding values from the literature or on our own.
One should note that the depth in \citealt{Prescott2012} is based on 1.1 arcsec diameter aperture (see section~2.1). 
We tentatively estimate the depths in the upcoming RIDEN survey by scaling the limiting magnitudes of the HCD-JF data ($u=27.1$ and $g=27.3$, Table~\ref{tab1}) to those in the Rubin/LSST co-added data ($u=26.1$ and $g=27.4$, see \citealt{Ivezic2019}).
As the Rubin/LSST survey coverage is out of range of the SDSS/BOSS quasar survey (e.g., \citealt{Lyke2020}), the number of quasars covered by RIDEN for each selection (Fig.~\ref{fig14}) is naively obtained by extending the number of quasars in the DR16Q catalogue \citep{Lyke2020} that satisfy our selection criteria (Section~\ref{s2}) over 9,376 deg$^2$ to the Rubin/LSST survey area (18,000 deg$^2$), which is estimated to be 215k and 229k in the $u$-band and $g$-band selection, respectively.

\begin{table}
\caption{
A summary of intensive Ly$\alpha$ nebula surveys.
}
\label{tab4}
\begin{tabular}{lccc}
    \hline
    Reference & SB$_\mathrm{Ly\alpha}$ limit & N$_\mathrm{qso}$ & Volume\\
    & (erg~s$^{-1}$cm$^{-2}$arcsec$^{-2}$) &  & (Gpc$^3$)\\
    \hline
    \multicolumn{4}{c}{IFU spectroscopy}\\
    \citealt{Borisova2016}        & $1.0\times10^{-18}$ &  17 & --- \\
    \citealt{ArrigoniBattaia2019} & $0.9\times10^{-18}$ &  61 & --- \\
    \citealt{Cai2019}             & $1.8\times10^{-18}$ &  16 & --- \\
    \citealt{Umehata2019}         & $0.3\times10^{-18}$ & --- & $1\times10^{-6}$\\
    \citealt{Fossati2021}         & $0.5\times10^{-18}$ &  27 & --- \\
    \citealt{Bacon2023}           & $3.0\times10^{-19}$ & --- & $3\times10^{-5}$\\
                                  & $0.7\times10^{-19}$ & --- & $3\times10^{-6}$\\
    \citealt{Martin2023}          & $0.5\times10^{-19}$ & --- & $2\times10^{-6}$\\
    \citealt{Tornotti2025}        & $0.5\times10^{-19}$ & --- & $4\times10^{-6}$\\
    \hline
    \multicolumn{4}{c}{Narrow-band selection}\\
    \citealt{Erb2011}             & $3.0\times10^{-18}$ & --- & $1\times10^{-4}$\\
    \citealt{Matsuda2011}         & $1.8\times10^{-18}$ & --- & $2\times10^{-3}$\\
    \citealt{Kikuta2019}          & $2.0\times10^{-18}$ & --- & $1\times10^{-3}$\\
    \citealt{Ramakrishnan2023}    & $4.4\times10^{-18}$ & --- & $9\times10^{-3}$\\
    \citealt{Li2024}              & $8.7\times10^{-18}$ &  52 & $1\times10^{-2}$\\
    \hline
    \multicolumn{4}{c}{Broad-band selection}\\
    \citealt{Prescott2012}        & $5.0\times10^{-17}$ & --- & $1\times10^{-1}$\\
    \citealt{Shimakawa2022}       & $1.4\times10^{-17}$ & 8683& ---\\
    This work ($u$-band)          & $0.5\times10^{-17}$ & 267 & $5\times10^{-2}$\\
    This work ($g$-band)          & $0.7\times10^{-17}$ & 216 & $1\times10^{-1}$\\
    RIDEN ($u$-band)              & $1.3\times10^{-17}$ & 215k& $1\times10^{2}$\\
    RIDEN ($g$-band)              & $0.7\times10^{-17}$ & 229k& $2\times10^{2}$\\
    \hline
\end{tabular}
\end{table}


\bsp	
\label{lastpage}
\end{document}